\documentclass[aps,prl,twocolumn,superscriptaddress,english]{revtex4}
\usepackage{amssymb}
\usepackage{graphicx}
\usepackage{amsmath}
\usepackage{amsthm}
\usepackage{amsfonts}
\usepackage[T1]{fontenc}
\usepackage[latin9]{inputenc}
\usepackage{array}
\usepackage{multirow}
\usepackage{color}
\usepackage{esint}
\usepackage{bm}
\usepackage{color}
\usepackage{bbm}
\usepackage{hyperref}
\usepackage{babel}
\usepackage{titlesec}

\begin{document}

\global\long\def\id{\mathbbm{1}}
\global\long\def\ui{\mathbbm{i}}
\global\long\def\ud{\mathrm{d}}

\title{Realization and detection of non-ergodic critical phases in optical Raman lattice}

\author{Yucheng Wang}
\affiliation{Shenzhen Institute for Quantum Science and Engineering, and Department of Physics,
Southern University of Science and Technology, Shenzhen 518055, China}
\affiliation{International Center for Quantum Materials, School of Physics, Peking University, Beijing 100871, China}
\affiliation{Collaborative Innovation Center of Quantum Matter, Beijing 100871, China}
\author{Long Zhang}
\affiliation{International Center for Quantum Materials, School of Physics, Peking University, Beijing 100871, China}
\affiliation{Collaborative Innovation Center of Quantum Matter, Beijing 100871, China}
\author{Sen Niu}
\affiliation{International Center for Quantum Materials, School of Physics, Peking University, Beijing 100871, China}
\affiliation{Collaborative Innovation Center of Quantum Matter, Beijing 100871, China}
\author{Dapeng Yu}
\affiliation{Shenzhen Institute for Quantum Science and Engineering, and Department of Physics,
Southern University of Science and Technology, Shenzhen 518055, China}
\author{Xiong-Jun Liu}
\thanks{Corresponding author: xiongjunliu@pku.edu.cn}
\affiliation{International Center for Quantum Materials, School of Physics, Peking University, Beijing 100871, China}
\affiliation{Collaborative Innovation Center of Quantum Matter, Beijing 100871, China}
\affiliation{Beijing Academy of Quantum Information Science, Xibeiwang East Rd, Beijing 100193, China}
\affiliation{CAS Center for Excellence in Topological Quantum Computation, University of Chinese Academy of Sciences, Beijing 100190, China}
\affiliation{Shenzhen Institute for Quantum Science and Engineering, and Department of Physics,
Southern University of Science and Technology, Shenzhen 518055, China}

\begin{abstract}
The critical phases, being delocalized but non-ergodic, are fundamental phases which are different from both the many-body localization and ergodic extended quantum phases, and have so far not been realized in experiment. Here we propose to realize such critical phases with and without interaction based on a topological optical Raman lattice scheme, which possesses one-dimensional spin-orbit coupling and an incommensurate Zeeman potential. We demonstrate the existence of both the noninteracting and many-body critical phases, which can coexist with the topological phase, and show that the critical-localization transition coincides with the topological phase boundary in noninteracting regime. The dynamical detection of the critical phases is proposed and studied in detail. Finally, we demonstrate how the proposed critical phases can be achieved based on the current cold atom experiments. This work paves the way to observe the novel critical phases.
\end{abstract}
\maketitle

{\em Introduction.---}Anderson localization is a fundamental and highly-explored quantum phenomenon in condensed matter physics~\cite{Anderson1958}, showing the disorder-induced localization of electronic wave-functions. The recent cold atom experiments have observed Anderson localization in the one-dimensional (1D) lattice with controlled disorder~\cite{Billy2008} and incommensurate quasiperiodic optical lattice~\cite{Roati2008}. The 1D disordered or incommensurate quantum systems can keep to be localized when interactions are considered, leading to the many-body localization (MBL), which is an ergodicity-breaking phase. The existence of the MBL phase has been well established in both theory~\cite{Huse2010,Huse2015,Altman2015,HuseX,Li,Lea1,Abanin2019} and experiment~\cite{Bloch1,Bordia,Bloch2,Bloch3,Bloch4}.

Between the localization and ergodic extended phases, a third type of fundamental phases, called critical phases, can exist without or with interactions, with the latter case leading to the many-body critical (MBC) phase which is an extended but nonthermal quantum many-body state~\cite{Wang2019}. Critical phases are important in understanding the transitions from localization or MBL to extended phases, and exhibit various interesting features, including the critical spectral statistics~\cite{Geisel1991,Fujita1986,GG2016}, multifractal behavior of wave-functions~\cite{Halsey1986,Mirlin2006,Dubertrand2014}, and dynamical evolutions~\cite{Abe1988,Geisel1997,Modugno2009}. However, the critical phases have not been realized in experiment. So far only few theoretical models may host critical phases, including the 1D extended Aubry-Andr\'{e}-Harper model with incommensurate off-diagonal hopping and on-site potential~\cite{Hatsugai1990,Takada2004,Chong2015} which further gives MBC phase in the interacting regime~\cite{Wang2019}, and the disordered 1D chain of $p$-wave superconductor~\cite{Cai2013,Hu2016,Wang2016}. However, these models are not realistic in experiment. Further, the experimental diagnostic for detecting the critical phases is lacking~\cite{Bloch4}.

In this letter we propose to realize critical phases in both single-particle and interacting regimes based on a 1D optical Raman lattice scheme with incommensurate Zeeman potential. The optical Raman lattice has been actively studied recently for realizing spin-orbit (SO) coupling and topological phases with ultracold atoms in both theory~\cite{LiuXJ2013,LiuXJ2014,Pan2015,Lepori2016,Zhou2017PRL,WangBZ2018,Pu2019,Zhao2019NJP,Zheng2019,Lu2019} and experiment~\cite{Liu2016,Sun2018,Song2018,Song2019}. The phase diagram with a broad region of critical phases is obtained, and the detection is further proposed and studied in detail. This proposal is of high feasibility and can be immediately achieved in the current experiments.

{\em Model.---}We consider a 1D SO coupled atomic quantum gas trapped in an optical Raman lattice~\cite{RamanReview2018} with an incommensurate Zeeman potential. The Hamiltonian
\begin{equation}
 H=H_0+U\sum_jn_{j\uparrow}n_{j\downarrow},
\label{Ham-z}
\end{equation}
where $U$ denotes the Hubbard interaction, the particle number operator $n_{i,\sigma}=c^{\dag}_{i,\sigma}c_{i,\sigma}$ at $i$-th site, with $c_{i,\sigma}$ the annihilation operator at spin $\sigma=\uparrow,\downarrow$, and
\begin{eqnarray}\label{ham-1}
H_0 &=& -t_0\sum_{\langle i,j\rangle}(c^{\dagger}_{i,\uparrow}c_{j,\uparrow}-c^{\dagger}_{i,\downarrow}c_{j,\downarrow})+\sum_{i}\delta_i(n_{i,\uparrow}-n_{i,\downarrow})\nonumber\\
 &+& \sum_i\bigr[t_{\rm so} (c^{\dagger}_{i,\uparrow}c_{i+1,\downarrow}-c^{\dagger}_{i,\uparrow}c_{i-1,\downarrow})+H.c.\bigr].
\end{eqnarray}
Here $t_0$ ($t_{\rm so}$) represents the spin-conserved (spin-flip) hopping strength between neighboring sites.
The Zeeman splitting $\delta_i = M_z \cos(2\pi\beta i+\phi)$ with an irrational number $\beta$
and a phase offset $\phi$ denotes the spin-dependent incommensurate potential with strength $M_z$, whose realization will be given later.
We note that the above model with a uniform Zeeman potential $\delta_i=m_z$  has been proposed and realized in optical Raman lattices and gives a 1D AIII class topological insulator~\cite{LiuXJ2013,WangBZ2018,Song2018,RamanReview2018}. For convenience, we set $t_0 = 1$ as the unit of energy and
$\beta=\frac{\sqrt{5}-1}{2}$, which is approached by
$\beta=\lim_{n \rightarrow \infty}\frac{F_{n-1}}{F_{n}}$. Here $F_{n}$ is the Fibonacci numbers defined recursively by  $F_{n+1}=F_{n-1}+F_{n}$, with $F_0=F_1=1$~\cite{Kohmoto1983}. In noninteracting cases, the rational approximation $\beta=F_{n-1}/F_{n}$ is taken for
system size $L=F_{n}$ to ensure a periodic boundary condition.

%%%%%%%%%%%%%%%%%%%%%%%%%%%%%%%%%%%%%%%%%%%%%%%
\begin{figure}
\centering
%\hspace*{-0.5cm}
\includegraphics[width=0.48\textwidth]{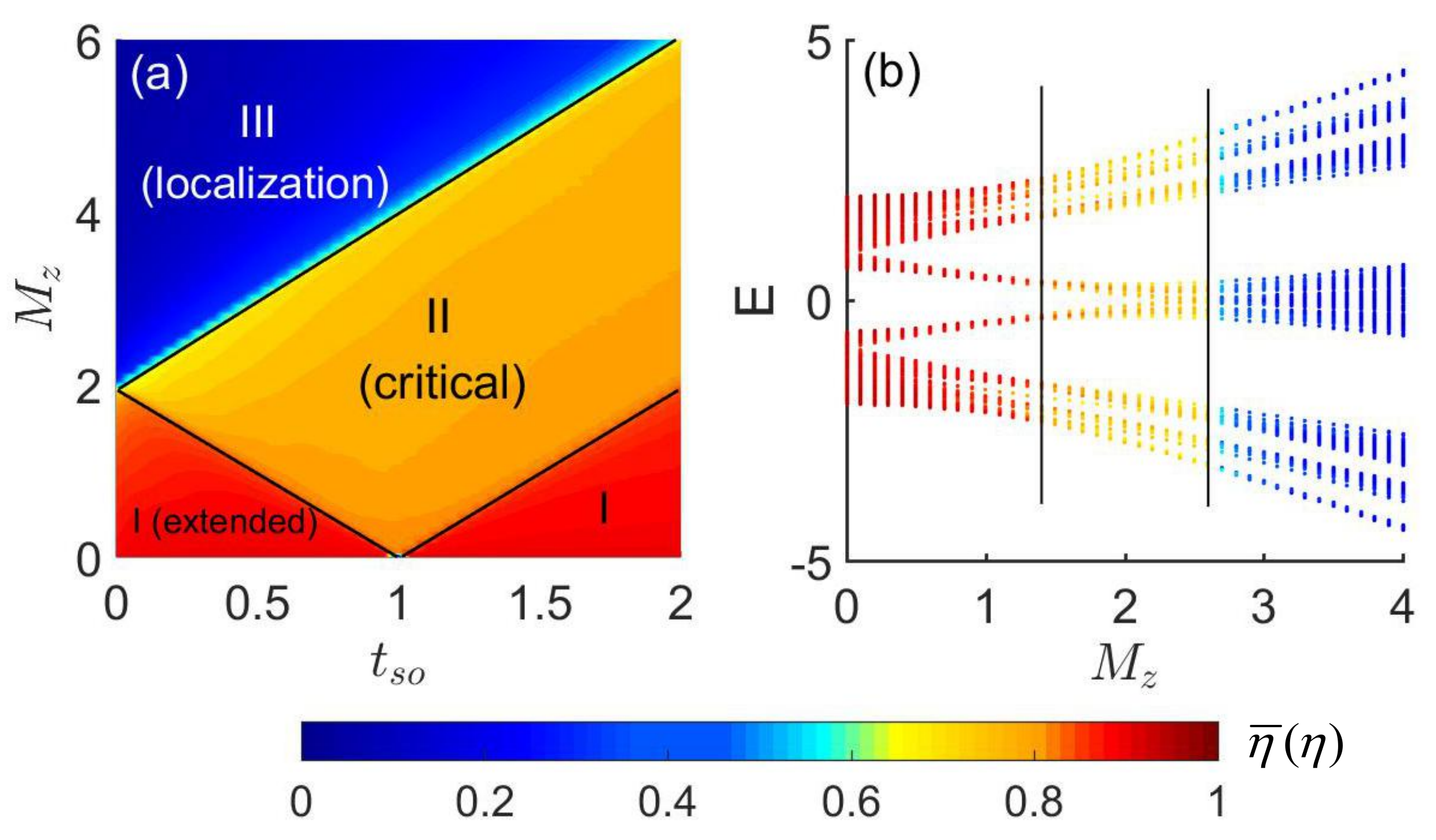}
%\par\end{raggedleft}
\caption{\label{01}
 (a) The mean FD $\overline{\eta}$ with $L=F_{14}=610$ as a function of incommensurate potential strength $M_z$ and $t_{\rm so}$. Region I, II and III correspond to extended, critical and localization phases respectively and the solid lines represent the boundaries of different phases. (b) FD $\eta$ of different eigenstates as a function of the corresponding energy eigenvalues and $M_z$ with size $L=610$ and $t_{\rm so}=0.3$. The two solid lines represent the phase boundaries $M_{z}^{c}=1.4$ and $2.6$.  Here we set $\phi=0$.}
\end{figure}
%%%%%%%%%%%%%%%%%%%%%%%%%%%%%%%%%%%%%%%%%%%%%%%%

{\em Noninteracting critical phase.---}We study first the noninteracting regime with $U=0$, which can be diagonalised exactly. In this regime, the phase diagram can be determined by the fractal dimension (FD), which for an arbitrary $m$-th eigenstate $|\psi_m\rangle=\sum_{j}^{L}[u_{m,j}c^{\dagger}_{j,\uparrow}+v_{m,j}c^{\dagger}_{j,\downarrow}]|vac\rangle$ is related to the inverse participation ratio (IPR) by $\eta=-\lim_{L\rightarrow\infty}\ln({\rm IPR})/\ln L$, with ${\rm IPR}(m) = \sum_{j} (u^4_{m,j}+v^4_{m,j})$. It is known that $\eta\to1$ ($0$) for the noninteracting extended (localized) states, while $0<\eta<1$ for critical states. To characterize the phases we define the mean IPR over all eigenstates: ${\rm MIPR}=(2L)^{-1}\sum_{m=1}^{2L}{\rm IPR}(m)$, and the mean FD being $\overline{\eta}=-\lim_{L\rightarrow\infty}\ln {\rm MIPR}/\ln L$. In Fig.~\ref{01} (a), we display $\overline{\eta}$ as a function of $M_z$ and $t_{\rm so}$, which clearly characterize three distinct phases, i.e., extended (I, with $\bar\eta\to1$), critical (II with $0<\bar\eta<1$), and localized (III, with $\bar\eta\to0$) phases.
The phase boundaries is precisely determined from a finite-size scaling analysis (see Supplemental Materials (SM)~\cite{SM}), given by
\begin{equation}
M_{z}^{c}=
\begin{cases}
2|t_0-t_{\rm so}|,\ \ \textrm{between I and II}, \\
2(t_0+t_{\rm so}),\ \textrm{between II and III}.
\end{cases}
\end{equation}
The FD $\eta$ also shows that all eigenstates in region II are critical [Fig.~\ref{01} (b)]. Note that
the critical-localized phase transition coincides the topological phase boundary. The in-gap edge states exist in the critical phase which is topological, but not in the localization phase. More details are presented in Supplementary Material~\cite{SM}.

%To further confirm that all states in region II are critical, i.e, $\overline{\eta}$ is fraction not originating from the coexistence of extended and localized states, we plot the FD $\eta$ of the entire energy spectrum in Fig.~\ref{01} (b). We see that $\eta$ are close to $1$, $0$ and fractions for all states on the left of the left solid line ($M_z^c=2(t_0-t_{\rm so})$), on the right of the right solid line ($M_z^c=2(t_0+t_{\rm so})$) and between the two lines respectively, which signals that the three phases from left to right
%are extended, critical and localized respectively. Furthermore, we find that critical-localization transition is also accompanied by the topological transition (see SM~\cite{SM}).
%%%%%%%%%%%%%%%%%%%%%%%%%%%%%%%%%%%%%%%%%%%%%%%
\begin{figure}
\centering
%\hspace*{-0.6cm}
\includegraphics[width=0.5\textwidth]{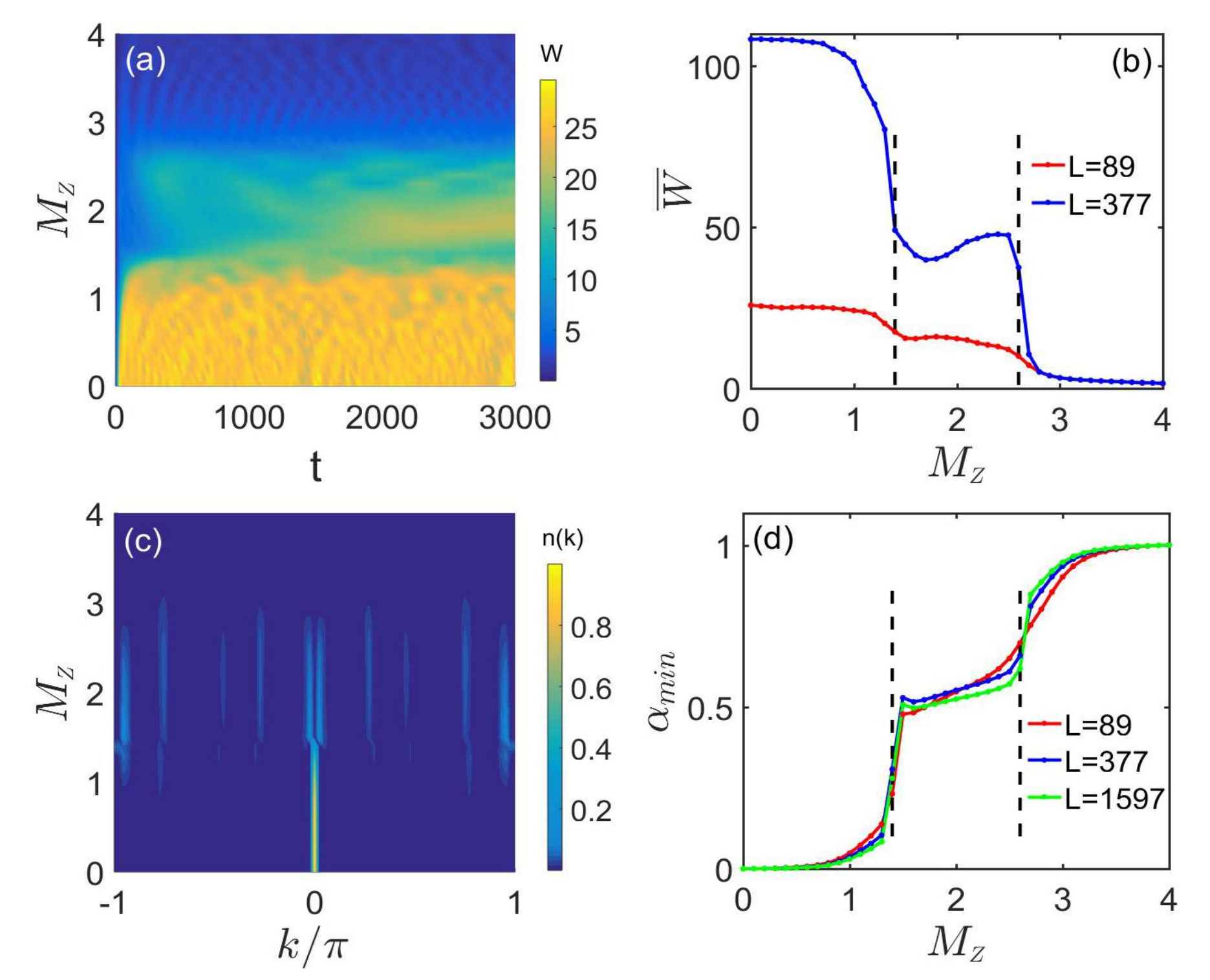}
\caption{\label{02}
  (a) $W$ as a function of $M_z$ and time $t$ for the system size $L=F_{10}=89$. (b) $\overline{W}$ versus $M_z$ for system size $L=89$ and $L=F_{13}=377$. (c) Ground states' momentum distributions $n(k)$ for different $M_z$ with fixed $L=89$. (d) $\alpha_{min}$ corresponding to the maximal momentum distribution of ground state versus $M_z$ for different sizes $L=89$, $L=377$ and $L=F_{16}=1597$. Here we take $t_{\rm so}=0.3$ and $\phi=0$.}
\end{figure}
%%%%%%%%%%%%%%%%%%%%%%%%%%%%%%%%%%%%%%%%%%%%%%%%

The realized phases can be experimentally detected by probing the dynamical evolution of spatial or momentum distributions of the quantum states. A useful quantity in characterizing the dynamical evolution of a wave packet~\cite{Billy2008,Roati2008} is the mean square displacement
\begin{eqnarray}
 W(t)=\biggr\{\sum_{j,\sigma=\uparrow,\downarrow}\bigr[j-(L+1)/2\bigr]^2\langle n_{j,\sigma}(t)\rangle\biggr\}^{1/2},
\end{eqnarray}
which is a measure of the width of the wave packet initially at $c^\dagger_{(L+1)/2,\downarrow}|0\rangle$ (let $L$ be odd).
Fig.~\ref{02} (a) displays how $W$ evolves with time (in units of $\hbar/t_0$) for different $M_z$.
We observe that for the extended phase obtained in the relatively small $M_z$, $W$ reaches a large and stable value in a fairly short time, while
for large $M_z$, $W$ remains very small all the time, signifying the localization of the wave packet. In comparison, for the critical phase,
$W$ grows gradually and slowly, different from both the extended and localization regimes.
The different behaviors can be quantified as
\begin{equation}
W(t)\sim t^{\kappa},
\end{equation}
where the dynamical index $\kappa$ can be shown to take $\kappa=1$, $\kappa=0$, and $\kappa\approx 1/2$
corresponding to the extended, localized, and critical phases, respectively (see SM~\cite{SM}).
A related observable to distinguish the phases is $\overline{W}=\frac{1}{N_t}\sum_{m=1}^{N_t}W(mT_c)$, which reflects the mean width of the wave packet over a long period of time. Here the sum is taken over a stroboscopic time evolution in steps of $T_c$. The result is shown in Fig.~\ref{02} (b) with $T_c=100$ and $N_t=30$. It is seen that $\overline{W}$ increases with the system size in both extended and critical phases, but remains a small size-independent value in the localized phase.
The two points where $\overline{W}$ decreases rapidly indicate the locations of phase boundaries, with an approximate plateau in-between marking the critical phase.

The detection of the critical phase can also be achieved from momentum distribution [see Fig.~\ref{02} (c)]
\begin{equation}
 n(k)=\frac{1}{L}\sum_{i,j=1}^{L}e^{-ik(i-j)}(\rho^{\uparrow}_{ij}+\rho^{\downarrow}_{ij}),
\end{equation}
where $\rho^{\sigma}_{ij}=\langle\psi_m|c_{i,\sigma}^{\dagger}c_{j,\sigma}|\psi_m\rangle$
are the single-particle density matrices with respect to the eigenstate $|\psi_m\rangle$. The momentum distribution exhibits localized, multifractal, and extended features in different regions. Quantitatively, we introduce the fractal index $\alpha(k)=-\ln n(k)/\ln L$, and examine the minimal index $\alpha_{\rm min}$ which characterizes the distribution peak $n_{\rm max}$. When $L\rightarrow\infty$, we see $\alpha_{\rm min}=0$, $\alpha_{\rm min}=1$, and $0<\alpha_{\rm min}<1$ for the extended, localization, and critical phases respectively [Fig.~\ref{02} (d)]. The experimental diagnostics for different phases are summarized in Table~\ref{wavefunction}.

\begin{table}\renewcommand{\arraystretch}{1.3}
  \centering
\begin{tabular}{|c|c|c|c|ll}
\cline{1-4}
    phases  &   extended & critical & localized & \\ \cline{1-4}
    $\kappa$(r) & 1 & $\approx 0.5$ & 0 & \\ \cline{1-4}
    $\overline{W}$(r)  & \ maximum\ &\ middle plateau\ &\ minimum\ &\\ \cline{1-4}
     $\alpha_{min}$(k) & 0 & $\in (0, 1)$ & 1 & \\ \cline{1-4}
\end{tabular}
\caption{The observables $\overline{W}$, $\kappa$ and $\alpha_{min}$ ($L\rightarrow\infty$) in different phases, where $r$ and $k$ imply the corresponding quantities measured in real space and momentum space, respectively.}\label{wavefunction}
\end{table}

%%%%%%%%%%%%%%%%%%%%%%%%%%%%%%%%%%%%%%%%%%%%%%%
\begin{figure}
\centering
%\hspace*{-0.7cm}
\includegraphics[width=0.5\textwidth]{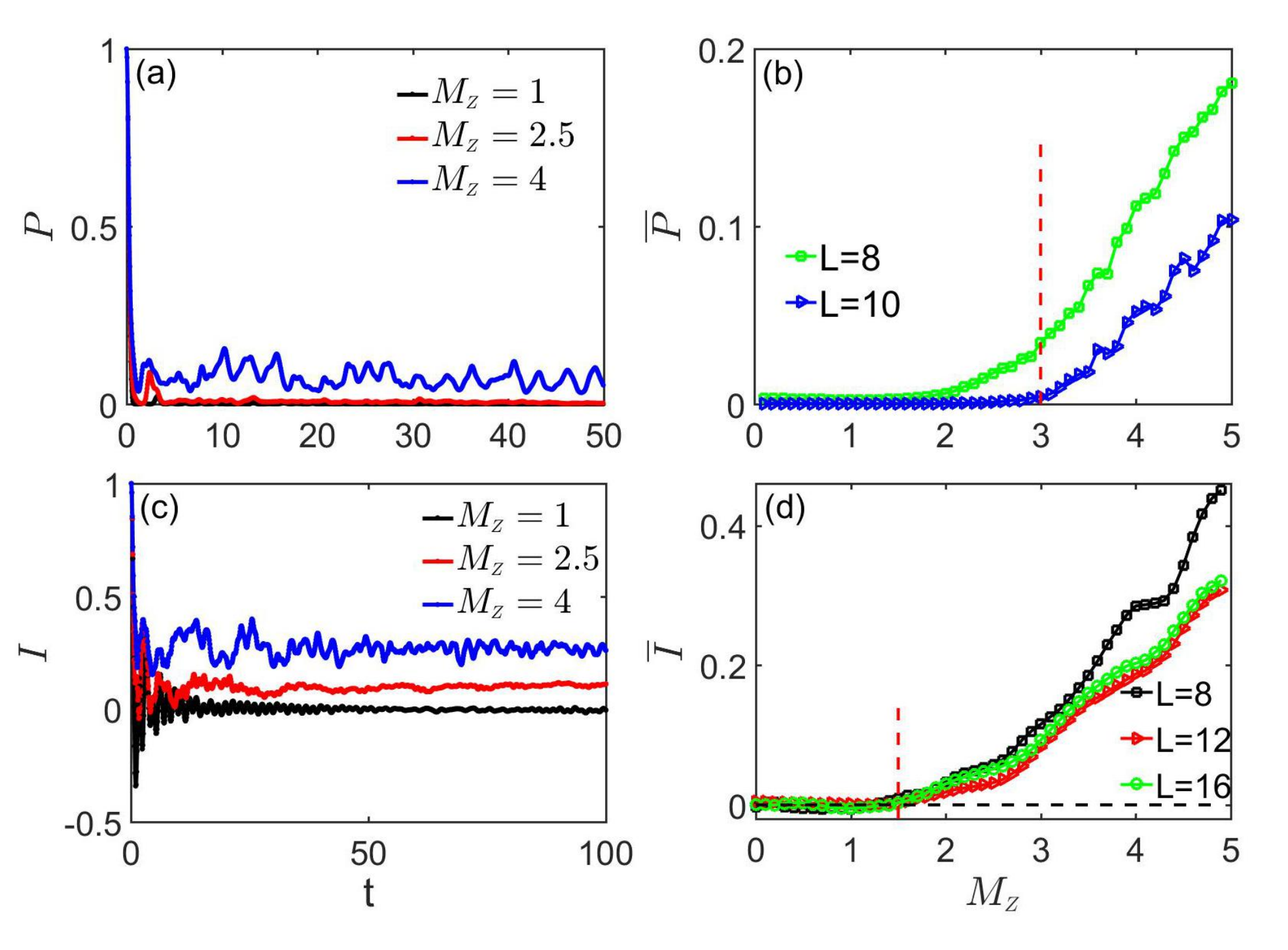}
\caption{\label{04}
  Time evolution of (a) the return probability with fixed $L=10$ and (c) the density imbalance $I$ with fixed $L=16$ for $M_z=1, 2.5, 4$, respectively. (b) $\overline{P}$ obtained from Eq.~(\ref{RP}) versus $M_z$ with system sizes $L=8, 10$ and the red dashed line denotes $M_z=3$. (d) $\overline{I}$ as a function of $M_z$ for different system sizes $L$. Red and black dashed lines correspond to $M_z=1.5$ and $\overline{I}=0$, respectively.  Here we fix $t_{\rm so}=0.3$ and use both the exact diagonalization method (for $L=8, 10$) and time-density matrix renormalization group method (for $L=12, 16$). To reduce the fluctuations, we take the average over $10$ realizations with different initial phase $\phi$.}
\end{figure}
%%%%%%%%%%%%%%%%%%%%%%%%%%%%%%%%%%%%%%%%%%%%%%%%

{\em Many-body critical phase.---} When the Hubbard interaction is included, the critical phase in the noninteracting regime may enter a new fundamental many-body state, namely, the MBC phase which is extended but not thermalizable~\cite{Wang2019}. We here propose to study the universal quench dynamics in the interacting regime, which on one hand confirms the existence of the MBC, one other hand provides experimental detection of the phase. For convenience, we take $U = 1$ and consider the quarter filling with the initial state $|\Psi(0)\rangle=|\uparrow0\uparrow0\uparrow0\uparrow\cdots\rangle$, namely, $N=L/2$, which is readily achievable in experiment.
We emphasize that our following results are independent of the specific configuration of the initial state.

We study the return probability of the many-body state under the evolution of the total Hamiltonian $H$, given by $P(t)=|\langle\Psi(0)|\Psi(t)\rangle|^2$, with $|\Psi(t)\rangle$ the state after evolution time $t$.
This observable has been measured in interacting many-body systems to observe the dynamical quantum phase transition~\cite{Roots2017,measureRP}.
By using $|\Psi(t)\rangle=\sum_{n}e^{-iE_nt}|\varphi_n\rangle\langle\varphi_n|\Psi(0)\rangle$, where $|\varphi_n\rangle$ are many-body eigenstates with eigenvalues $E_n$, one can obtain the long-time average of the return probability
\begin{equation}
 \overline{P}=\lim_{T \rightarrow \infty}\frac{1}{T}\int_0^{T}dt\,P(t)=\sum_{n}|\langle\varphi_n|\Psi(0)\rangle|^4,
\label{RP}
\end{equation}
which resembles the IPR with the leading term $\overline{P}\sim D_H^{-\eta}$. Here $D_H$ is the Hilbert space size and $\eta\to1$, $0<\eta<1$ and $\eta\to0$ for the ergodic, MBC, and MBL phases respectively~\cite{Wang2019}.
This result tells that the MBL (ergodic extended) phase preserves (losses) the local quantum information of initial state after a long time evolution~\cite{Luitz}. The critical phase losses local information with a lower speed.
To further distinguish MBC from ergodic extended phase, we consider the density imbalance
\begin{equation}
I=\frac{N_{\rm odd}-N_{\rm even}}{N_{\rm odd}+N_{\rm even}},
\end{equation}
where $N_{\rm odd}$ ($N_{\rm even}$) denotes the atom number on odd (even) sites. The long-time average of particle number at site $j$ can be predicted by a microcanonical ensemble analysis if the system is ergodic~\cite{Rigol2008}, i.e.,
$ \overline{n}_j=\lim_{T \rightarrow \infty}\frac{1}{T}\int_0^{T}dt\,n_j(t)=\langle n_j\rangle_{\rm microcan}(E_0)$,
where $E_0$ is the energy of the initial state. Thus the particle distribution of $|\Psi(t)\rangle$ for large $t$ is independent of the initial distribution, hence the final distribution is uniform, giving $I=0$. In contrast, $I$ remains a finite value after a long time evolution for the thermal critical phase. Therefore, the MBC phase can be distinguished from the other two through a combined measurement of $P$ and $I$.

Fig.~\ref{04} (a) shows the behavior of the return probability $P$, which
quickly decays to zero when $M_z=1$ or $2.5$, signifying that the phases are extended.
In contrast, $P$ maintains a nonvanishing value during the evolution at $M_z=4$, indicating that the system enters the MBL phase.
The long-time average $\overline{P}$ is displayed as a function of $M_z$ in
Fig.~\ref{04} (b).
One can see that $\overline{P}$ increases from zero to non-zero around the point $M_z=3$ for $L=10$. However, for $L=8$, the transition point moves to the regime $M_z<3$. We attributes it to the finite-size effect~\cite{reason}.
Hence we can conclude that when $L\rightarrow\infty$, the critical point of the delocalization-localization transition is a bit larger than $M_z=3$.
We then examine the density imbalance $I$, as shown in Fig.~\ref{04} (c). It demonstrates that the system is ergodic when $M_z=1$, but non-ergodic when $M_z=2.5$ or $4$.
Combing Fig.~\ref{04} (a), we find the system to be in the MBC phase for $M_z=2.5$.
Fig.~\ref{04} (d) shows the long-time average imbalance $\overline{I}$, which is numerically calculated by using $\overline{I}=\frac{2}{T}\int_{T/2}^{T}dtI(t)$ with $T=100$,% and the time interval $dt=0.1$,
and it approximates the stable value of $I$ after long time evolution. We see that the ergodicity breaking transition occurs near $M_z=1.5$, insensitive to system size. The combined measurements show that the approximate region with $M_z\in (1.5, 3)$ belongs to the extended and nonthermal MBC phase .

\begin{figure}
\includegraphics[width=0.48\textwidth]{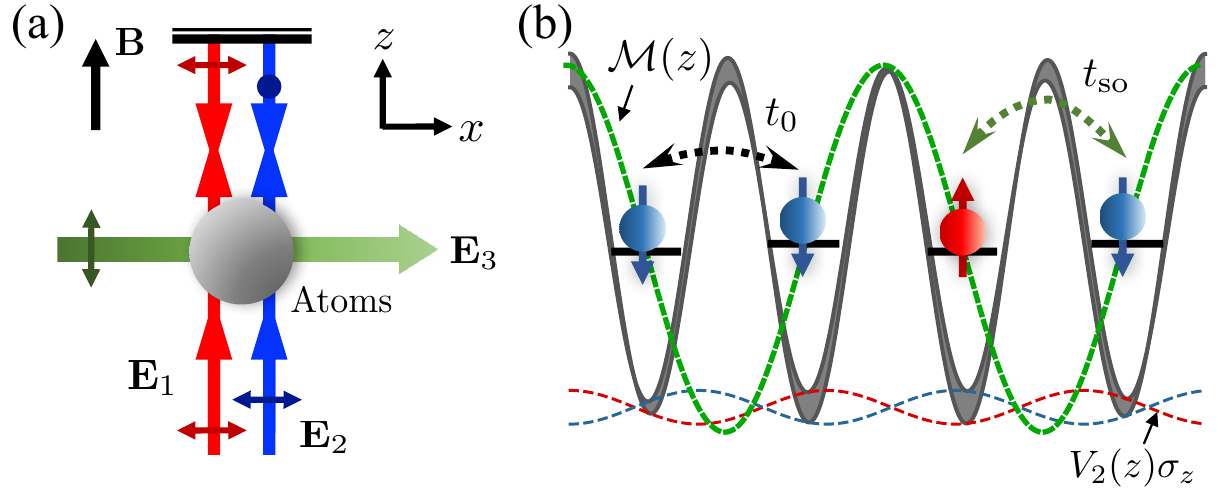}
\caption{Experimental realization in cold atoms. (a) Schematic of experimental setup. A standing wave ${\bf E}_1$ with $x$ polarization generates the spin-independent main lattice.
Another beam ${\bf E}_2$, formed by two counter-propagating lights with mutually perpendicular polarization, gives a spin-dependent incommensurate lattice.
A plane wave ${\bf E}_{3}$ is applied along $x$ direction to form a Raman coupling potential with one component of ${\bf E}_{1}$.
(b) The spin-conserved ($t_0$) and -flipped ($t_{\rm so}$) hoppings are induced by the main lattice and Raman potentials, respectively. The secondary lattice
provides an incommensurate spin-dependent offset.
}\label{fig_setup}
\end{figure}

{\em Proposal for experimental realization.---}
We propose a highly feasible experimental setup to realize the Hamiltonian (\ref{Ham-z})
based on optical Raman lattices~\cite{LiuXJ2013,Song2018,WangBZ2018}.
As depicted in Fig.~\ref{fig_setup} (a), a standing-wave beam  ${\bf E}_{1}$ (red) with $x$ polarization
and a plane wave ${\bf E}_{3}$ (green) with $z$ polarization are applied to generate a spin-independent main lattice $V_1(z)$
and a Raman coupling potential ${\cal M}(z)$ simultaneously. The former induces the spin-conserved hopping and the latter produces the spin-flip hopping [Fig.~\ref{fig_setup} (b)].
In addition, another standing wave ${\bf E}_{2}$ (blue),
formed by two counter-propagating lights with mutually perpendicular polarization, is used to produce
a spin-dependent lattice $V_2(z)$~\cite{WangBZ2018,Mandel2003}, which provides a secondary incommensurate Zeeman potential.

With the bias field ${\bf B}$ applied along the $z$ direction, the standing wave ${\bf E}_1$ with $x$ polarization can be written as
${\bf E}_1=\hat{e}_+E_{1+}\cos(k_1z)+\hat{e}_-E_{1-}\cos(k_1z)$ with $E_{1\pm}$ being the ampitudes.
For alkali-metal atoms, the optical lattice potential is spin independent,
which reads $V_1(z)=V_{\rm m}\cos^2({k}_1z)$ with the depth $V_{\rm m}\propto (E^2_{1+}+E^2_{1-})/\Delta$, and creates a tight-binding primary lattice.
Here $\Delta$ denotes the coupling detuning.
Having the plane-wave field ${\bf E}_3=\hat{e}_zE_3 e^{\ui k_3x}$,
one Raman potential ${\cal M}(z)=M_{\rm R}\cos({k}_1z)$, with $M_{\rm R}\propto E_{1-}E_3/\Delta$, can be
generated by $E_3$ and $E_{1-}$ components.
The Raman and lattice potentials
satisfy a relative antisymmetry, which ensures a spin-flipped
hopping along the $z$ direction~\cite{LiuXJ2013,RamanReview2018}.
The other standing-wave field reads ${\bf E}_2=\hat{e}_+E_{2+}\cos(k_2z)+\hat{e}_-E_{2-}\sin(k_2z)$
(see Refs.~\cite{WangBZ2018,Mandel2003,SM}).
The secondary lattice $V_2(z)\sigma_z=V_{\rm s}\cos^2({k}_2z)\sigma_z$, with $V_{\rm s}\propto (E^2_{2+}-E^2_{2-})/\Delta$, provides
a weak spin-dependent energy offset.
When $V_{\rm s}/V_{\rm m}\ll 1$,
this incommensurate lattice can be regarded as a perturbation such that
the spin-conserved and -flipped hoppings can be only determined by the primary lattice and Raman potentials [Fig.~\ref{fig_setup}(b)].
In our setting, the irrational number $\beta=k_2/k_1$ with the lattice wavenumbers $k_{1,2}$ being easily tunable in experiment~\cite{SM}.
Take $^{40}$K atoms as an example. We choose  $|\uparrow\rangle=|F=7/2, m_F=+7/2\rangle$ and $|\downarrow\rangle=|9/2, +9/2\rangle$ to construct the spin-$1/2$ system,
and set the wavelengths of ${\bf E}_{1,2}$ to be $\lambda_1=$768nm and $\lambda_2=$780nm, which yields $\beta\approx0.9846$.
We then have $V_{\rm s}/V_{\rm m}\sim E_2^2/27E_1^2$ and $M_{\rm R}/V_{\rm m}\sim E_3/E_1$~\cite{SM}.
One can tune the lattice and Raman potentials freely by the laser intensity.
For example, the settings $V_{\rm m}=4.2E_{\rm r}$ and $M_{\rm R}=1.3E_{\rm r}$ with $E_r\equiv\hbar^2k_1^2/2m$ give $t_{\rm so}\simeq t_0$; the critical phase corresponds to
the parameter region $|M_z|<4t_0$ according to Fig.~\ref{01}(a). This region can be easily achieved by tuning $V_{\rm s}$ in the range from 0 to less than 1.25 $E_{\rm r}$.

{\em Conclusion.---}We have proposed a highly feasible 1D SO coupled model with incommensurate Zeeman potential for realizing critical phases in a broad phase diagram region separating from the extended and localized phases. In the noninteracting regime, we showed that the critical phase can be detected by measuring the mean square displacement of the wave packet after a fixed evolution time in real space or measuring the momentum distributions. With interactions, we proposed two observables, i.e., the return probability of the initial state and the density imbalance, to distinguish the MBC phase from both the ergodic and MBL phases. This work opens a broad avenue with high experimental feasibility to explore critical phases in ultracold atoms.

We thank Immanuel Bloch for helpful comments. This work was supported by National Nature Science Foundation of China (11825401, 11761161003, and 11921005), the National Key R\&D Program of China (2016YFA0301604), Guangdong Innovative and Entrepreneurial Research Team Program (No.2016ZT06D348), the Science, Technology and Innovation Commission of Shenzhen Municipality (KYTDPT20181011104202253), and the Strategic Priority Research Program of Chinese Academy of Science (Grant No. XDB28000000).
%
%%%%%%%%%%%%%%%%%%%%%%%%%%%%%%%%%%%%%%%%%%%%%%

\onecolumngrid

\newpage{}

\section*{\large Supplementary Material:\\Realization and detection of non-ergodic critical phases in optical Raman lattice}

\setcounter{equation}{0} \setcounter{figure}{0} \setcounter{table}{0}
\setcounter{page}{1} \makeatletter \global\long\def\theequation{S\arabic{equation}}
 \global\long\def\thefigure{S\arabic{figure}}
 \global\long\def\thebibitem{S\arabic{reference}}

This Supplementary Materials provides the details of the finite size scaling analysis, multifractal analysis, the topological phase transitions, diffusion dynamics, and the experimental realization of the present model.
\section{I. Finite size scaling}
For convenience, we rewrite the Hamiltonian discussed in the main text:
\begin{eqnarray}\label{ham-F1}
H_0 &=& -t_0\sum_{\langle i,j\rangle}(c^{\dagger}_{i,\uparrow}c_{j,\uparrow}-c^{\dagger}_{i,\downarrow}c_{j,\downarrow})+\sum_{j}M_z\cos(2\pi\beta j+\phi)(n_{j,\uparrow}-n_{j,\downarrow})\nonumber\\
 &+& \sum_i\bigr[t_{so} (c^{\dagger}_{i,\uparrow}c_{i+1,\downarrow}-c^{\dagger}_{i,\uparrow}c_{i-1,\downarrow})+H.c.\bigr],
\end{eqnarray}
For eigenvalue $E_m$, the corresponding eigenstate can be written as $|\Psi_m\rangle=\sum_{j}^{L}[u_{m,j}c^{\dagger}_{j,\uparrow}+v_{m,j}c^{\dagger}_{j,\downarrow}]|vac\rangle$. Using the equation $H|\Psi_m\rangle=E_m|\Psi_m\rangle$, one can obtain the following explicit form
 \begin{eqnarray}
 -t_0(u_{m,j+1}+u_{m,j-1})+\delta_ju_{m,j}+t_{so}(v_{m,j+1}-v_{m,j-1})&=&E_mu_{m,j},\label{tb3}\\
 t_0(v_{m,j+1}+v_{m,j-1})-\delta_jv_{m,j}+t_{so}(u_{m,j-1}-u_{m,j+1})&=&E_mv_{m,j},\label{tb4}
\end{eqnarray}
where $\delta_j = M_z\cos(2\pi\beta j+\phi)$. If we introduce a vector $|\Psi_m\rangle=[u_1, v_1, u_2, v_2,\cdots,u_L, v_L]^{T}$, solving Eq.(\ref{tb3}) and
Eq.(\ref{tb4}) reduces to an eigenvalue problem of a $2L\times 2L$ matrix. Then we can calculate the eigenvalues and corresponding eigenstates of the
Hamiltonian (\ref{ham-F1}). When the irrational number $\beta$ is approached by $\beta=\frac{F_{n-1}}{F_{n}}$, where $F_{n}$ is the Fibonacci numbers as discussed in the main text, Eq.(\ref{tb3}) and Eq.(\ref{tb4}) is periodic with period $L=F_{n}$. From Bloch's theorem, we have ${{u_{j+L}}\choose{v_{j+L}}}=e^{ikL}{{u_{j}}\choose{v_{j}}}$ ($j=1, 2, \cdots, L$), with $k\in [-\frac{\pi}{L}, \frac{\pi}{L})$. Unless otherwise stated, we take $k= 0$ when using periodic boundary conditions (PBC).

To obtain the phase boundaries of this model, we perform a finite-size scaling analysis. Since the proposed model has a pure energy spectrum, i.e., there don't exist mobility edges, we can define a mean participation ratio: $I_L=\frac{1}{2L}\sum_{n=1}^{2L}\frac{1}{\sum_{i=1}^{L} (u^4_{n,i}+v^4_{n,i})}$. We further define:
$\sigma_L=(I_L/L)^{1/2}$, which tends to $0$ when $L\rightarrow\infty$ in the localized phase and is approximate to $1$ in the extended phase, so it can be used as the order parameter. Near the transition points, three critical exponents can be introduced~\cite{Hashimoto,Wang2}:
\begin{equation}\label{exponent}
\xi \sim |\Delta M_z|^{-\nu},\qquad
I \sim (\Delta M_z)^{-\gamma},\qquad
\sigma \sim (\Delta M_z)^{\beta},
\end{equation}
where $I=\lim_{L\rightarrow\infty}I_L$ and $\sigma=\lim_{L\rightarrow\infty}\sigma_L$, $\xi$ is the correlation or localization length, $\Delta M_z=(M_z-M^c_{z})/M^c_{z}$ with $M^c_{z}$ being the transition points. Near $M^c_{z}$, we assume a finite size scaling relationship when this system's size is finite:
\begin{equation}\label{scaling}
\sigma_L^2L^{1-\gamma/\nu}=f(L^{1/\nu}(\Delta M_z)),
\end{equation}
where $f(x)$ is the scaling function \cite{Hashimoto,Wang2}.

\begin{figure}
\includegraphics[width=0.6\textwidth]{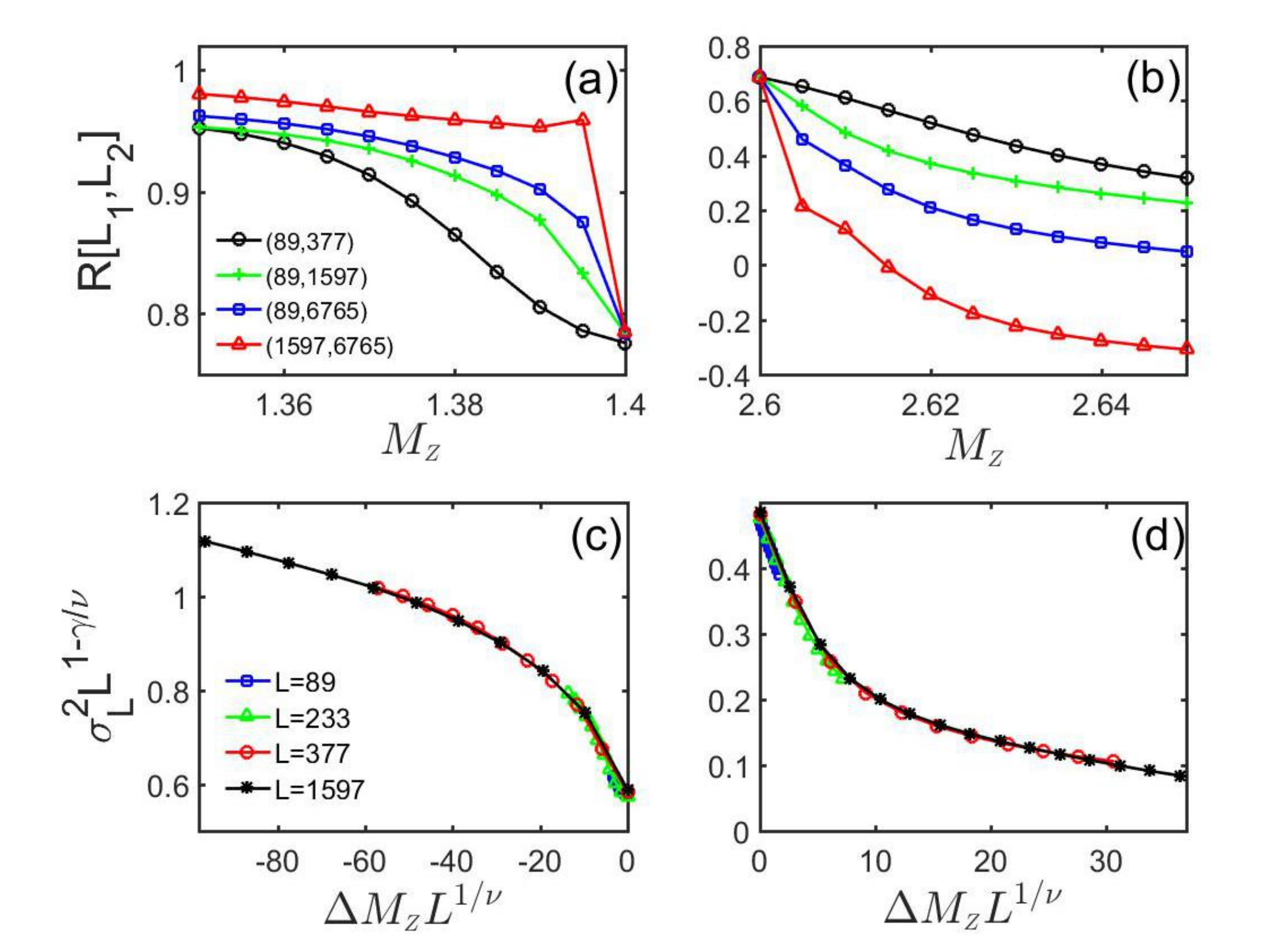}
\caption{\label{S3}
  $R[L_1,L_2]$ versus $M_z$ for several pairs of $(L_1,L_2)$ corresponding to the transition from the extended phase to critical phase in (a) and the transition
from the critical phase to localized phase in (b). $\sigma_L^2L^{1-\gamma/\nu}$ as a function of $\Delta M_z L^{1/\nu}$ for different $L$ with (c) $M_z\leq 2|t_0-t_{so}|$ and (d) $M_z\geq 2(t_0+t_{so})$. Different lines are superposed together with setting $\nu=1$. Here we fix $t_0=1$ and $t_{so}=0.3$ and take PBC.}
\end{figure}

Since $\Delta M_z=0$ at transition points, then Eq.~(\ref{scaling}) becomes $\sigma_L^2=f(0)L^{\gamma/\nu-1}$. For different sizes $L_1$ and $L_2$, a function of two size-variables can be defined as:
\begin{equation}
R[L_1,L_2]=\frac{\ln(\sigma_{L_1}^2/\sigma_{L_2}^2)}{\ln(L_1/L_2)}+1,
\end{equation}
which equals to $\gamma/\nu$ at transition points for any pair $(L_1, L_2)$. Fig. \ref{S3} (a) and (b) display the behaviors of $R[L_1,L_2]$ with increasing
$M_z$. We can determine two transition points $V_c=1.4$ (corresponding the transition from the extended phase to critical phase) and $V_c=2.6$ (corresponding the transition from the critical phase to localized phase) from the crossing point, which correspond to $M^c_{z}=2|t_0-t_{so}|$ and $M^c_{z}=2(t_0+t_{so})$ respectively, and the corresponding critical exponents are about $\gamma/\nu=0.78$ and $\gamma/\nu=0.68$.
The critical exponent $\nu$ can be obtained by calculating $\sigma_L^2L^{1-\gamma/\nu}$ as a function of $L^{1/\nu}(\Delta M_z)$ and making them superpose together for different $L$ as shown in Fig.~\ref{S3} (c) and (d). We see that lines corresponding to different $L$ superpose
together if setting $\nu=1$, which indicates $\nu=1$ at both the transition points from the extended to critical phases and from the critical to localized phases.
%%%%%%%%%%%%%%%%%%%%%%%%%%%%%%%%%%%%%%%%%%%%%%%
%%%%%%%%%%%%%%%%%%%%%%%%%%%%%%%%%%%%%%%%%%%%%%%%

%%%%%%%%%%%%%%%%%%%%%%%%%%%%%%%%%%%%%%%%%%%%%%%
\begin{figure}[h]
\centering
\includegraphics[width=0.95\textwidth]{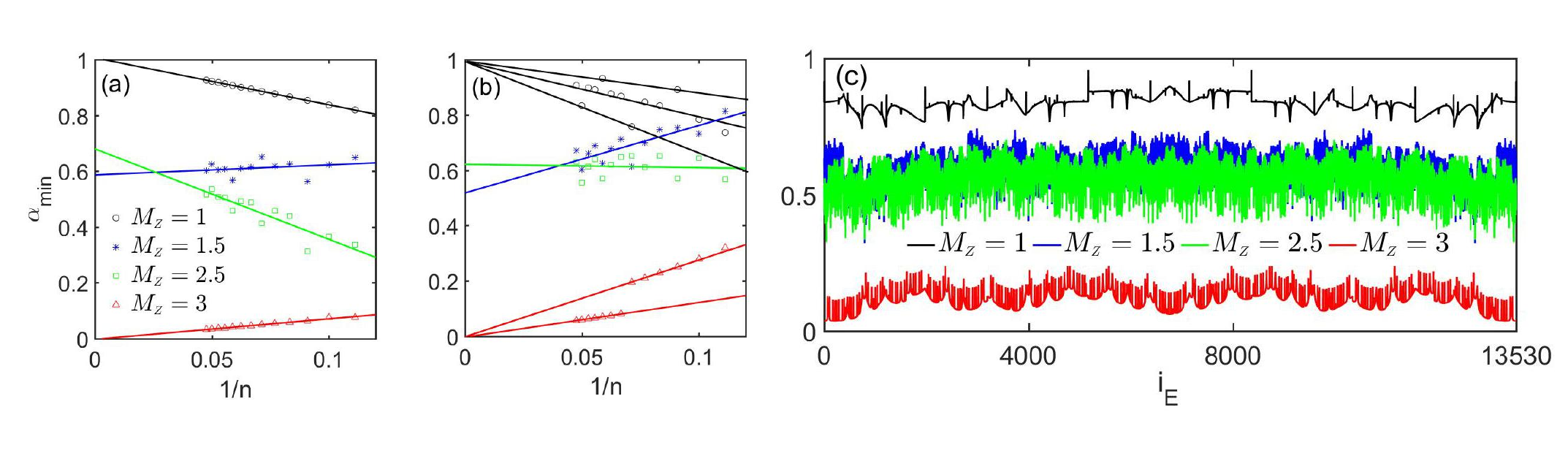}
\caption{\label{S1}
$\alpha_{min}$ as a function of $1/n$ for the lowest state (a) and the center state (b) of the spectrum for different incommensurate potential strengths $M_z$. (c) $\alpha_{min}$ versus spectrum index $i_E$ with fixed $L=F_{19}=6765$ and the number of energy mode
is $2L=13530$. Here we take PBC and fix $t_0=1$ and $t_{so}=0.3$, so the critical range is $M_z\in [1.4, 2.6]$.}
\end{figure}
%%%%%%%%%%%%%%%%%%%%%%%%%%%%%%%%%%%%%%%%%%%%%%%%

\section{II. Multifractal analysis}
To further confirm that all eigenstates in the proposed critical phase are critical, we investigate the scaling behavior of wave-functions. The probability measure at the lattice site $j$ is $p_j=n_{j, \uparrow}+n_{j, \downarrow}$ for a normalized wave-function and the corresponding scaling index $\alpha_j$ is defined by $p_j=L^{-\alpha_j}$. For extended states, all the lattice sites satisfy $\alpha_j\rightarrow 1$ when $L\rightarrow\infty$. For localized states, there exist non-vanishing probabilities only on a finite number of
sites, i.e., $\alpha_j\rightarrow0$ for these sites but $\alpha_j\rightarrow\infty$ for other sites. For critical states, $\alpha_j$ has a distribution, i.e., $\alpha_j\in [\alpha_{min}, \alpha_{max}]$ with $0 < \alpha_{min} <1$. Therefore, $\alpha_{min}$ when $L\rightarrow\infty$ can be used to identify extended ($\alpha_{min}=1$), critical ($0<\alpha_{min}<1$) and localized ($\alpha_{min}=0$) states.
Fig. \ref{S1} (a) and (b) present the $\alpha_{min}$ of two typical eigenstates at the lowest and the center of the spectrum respectively. We see that $\alpha_{min}$ of the two eigenstates tend to $1$ for $M_z=1$ (in the extended phase), to $0$ for $M_z=3$ (in the localized phase), and
to the values far from $0$ and $1$ for $M_z=1.5$ and $M_z=2.5$ (in the critical phase).
Fig. \ref{S1} (c) displays the $\alpha_{min}$ of all the eigenstates with system size $L=F_{19}=6765$.
There are no dramatic changes of $\alpha_{min}$ with increasing spectrum index $i_E$ for all incommensurate potential strengths $M_z$, which suggests
that there doesn't exist mobility edge in the energy spectrum. Therefore, all wave-functions are extended in the region I, critical in the region II, and
localized in the region III of Fig. 1 (a) in the main text.

%%%%%%%%%%%%%%%%%%%%%%%%%%%%%%%%%%%%%%%%%%%%%%%
\begin{figure}
\includegraphics[width=0.95\textwidth]{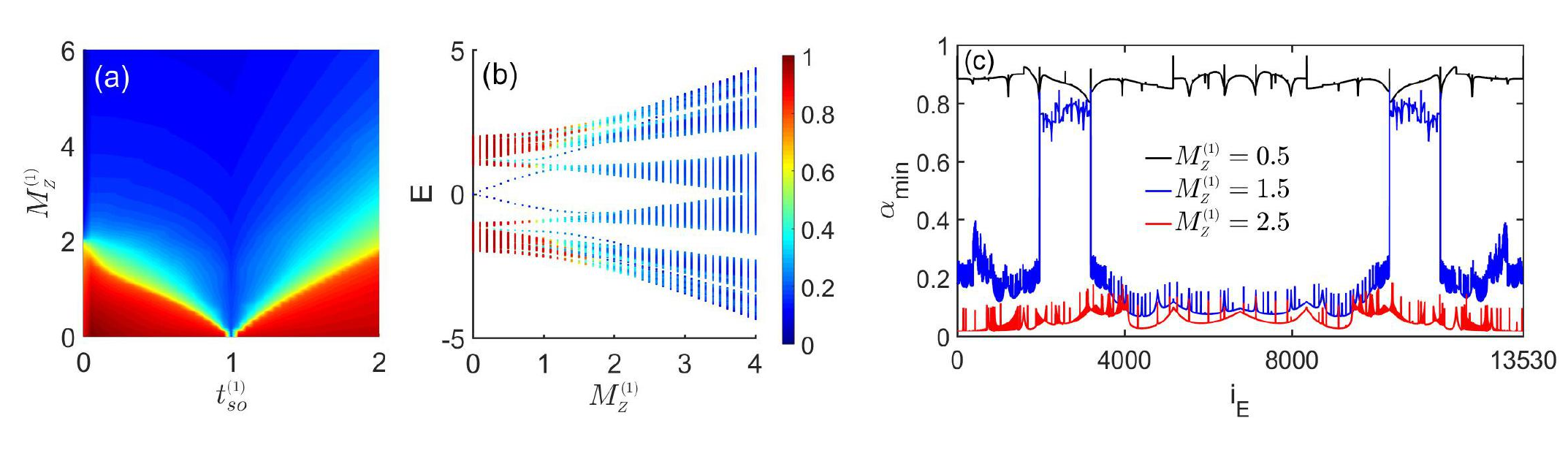}
\caption{\label{S4n}
  (a) Mean fractal dimension $\overline{\eta}$ with $L=500$ as a function of $M^{(1)}_z$ and $t^{(1)}_{so}$. (b) Fractal dimension $\eta$ of different eigenstates as a function of the corresponding energy eigenvalues and incommensurate potential strength $M^{(1)}_z$ with size $L=500$. (c) $\alpha_{min}$ versus spectrum index $i_E$. The system size is $L=6765$ and the number of energy mode is $2L=13530$. Here we fix $t^{(1)}_{0}=1$ and $t^{(1)}_{so}=0.5$ for (b) and (c).}
\end{figure}
%%%%%%%%%%%%%%%%%%%%%%%%%%%%%%%%%%%%%%%%%%%%%%%%

\section{III. Absence of critical phase for spin-independent incommensurate potential}

For comparison, we replace the incommensurate Zeeman potential of the Hamiltonian (\ref{ham-F1}) with a spin-independence incommensurate potential. We shall see that no critical phase can be obtained. The Hamiltonian reads
\begin{eqnarray}\label{ham-M1}
H_{1} &=& -t^{(1)}_0\sum_{\langle i,j\rangle}(c^{\dagger}_{i,\uparrow}c_{j,\uparrow}-c^{\dagger}_{i,\downarrow}c_{j,\downarrow})
+\sum_{i}\delta^{(1)}_i(n_{i,\uparrow}+n_{i,\downarrow}) \nonumber\\
 &&+ \sum_i\bigr[t^{(1)}_{so} (c^{\dagger}_{i,\uparrow}c_{i+1,\downarrow}-c^{\dagger}_{i,\uparrow}c_{i-1,\downarrow})+H.c.\bigr],
\end{eqnarray}
where the incommensurate potential $\delta^{(1)}_i = M^{(1)}_z cos(2\pi\beta i+\phi)$ is same for spin up and spin down particle.

According to the main text, the critical state exhibits multifractal structure of the wave-functions, giving the mean fractal dimension $0<\eta (\overline{\eta})<1$. In Fig.~\ref{S4n} (a), we show the mean fractal dimension $\overline{\eta}$ as a function of $M^{(1)}_z$ and $t^{(1)}_{so}$ with fixed $t^{(1)}_0=1$ and also find a region where $\overline{\eta}$ is far from $1$ and $0$. To see it clearly, we plot energy eigenvalues and the fractal dimension $\eta$ of the corresponding eigenstates as a function of potential strength $M^{(1)}_z$ in Fig.~\ref{S4n} (b) and we see the region that $0<\overline{\eta}<1$ corresponding to a intermediate regime where extended and localized states coexistent. In Fig.~\ref{S4n} (c), we show  $\alpha_{min}$ of all the eigenstates with system size $L=F_{19}=6765$. For $M^{(1)}_z=0.5$, all states are extended and for $M^{(1)}_z=2.5$, all states are localized. For $M^{(1)}_z=1.5$, there exist dramatic changes of $\alpha_{min}$ with increasing spectrum index $i_E$, suggesting mobility edge existing in the energy spectrum of the intermediate regime. However, we do not have that all states are critical. Thus no critical phase is obtained for this system.

\section{IV. Topological phase transition}

%%%%%%%%%%%%%%%%%%%%%%%%%%%%%%%%%%%%%%%%%%%%%%%
\begin{figure}
\centering
\includegraphics[width=0.65\textwidth]{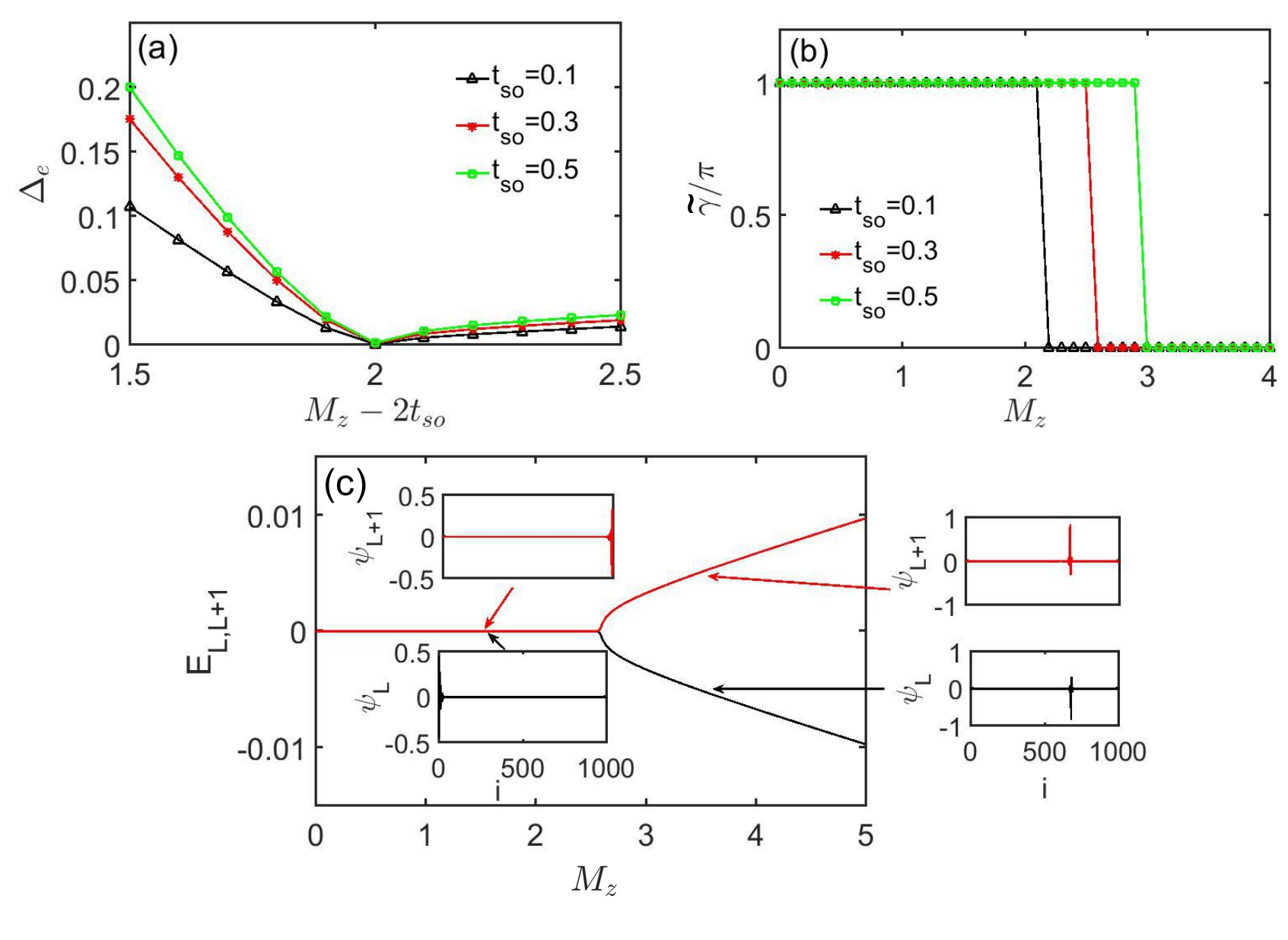}
\caption{\label{S6}
(a) The energy gap $\Delta_{e}$ as a function of $M_z-2t_{so}$ for the system size $L = F_{12}= 233$. (b) The numerical calculated Berry phase versus $M_z$ for different $t_{so}$ for the system size $L = F_{10}= 89$. (c) Eigenvalues $E_L$ and $E_{L+1}$, where the eigenvalues have been listed in ascending order, and the corresponding wave functions for $M_z=1.5$ and $M_z=3.5$ with fixed $t_{so}=0.3$ and $L=500$ for the Hamiltonian (\ref{ham-F1}). Here we take PBC for (a) and (b) and OBC for (c).}
\end{figure}
%%%%%%%%%%%%%%%%%%%%%%%%%%%%%%%%%%%%%%%%%%%%%%%%
If the Zeeman potential in Hamiltonian (\ref{ham-F1}) is uniform, i.e., $\delta_i=\delta$, this system is topological when $|\delta| < 2t_0$ and is topologically trivial when $|\delta| > 2t_0$~\cite{Liu2013}. We below discuss the topological transitions of this system at half filling when adding the spin-dependent incommensurate potential. The eigenvalues of this system under PBC are arranged in ascending order and we consider the gap given by $\Delta_{e}=E_{L+1}-E_{L}$. Fig.~\ref{S6} (a) shows the gap for systems with different $t_{so}$ and we see that the gap closes at the critical-localization transition point $M^c_z=2(t_0+t_{so})$ and there exist nonzero gaps when $M_z$ is smaller than or exceeds the transition point $M^c_z$. Thus this system may exist topological transition at the critical-localization transition point. To characterize the topological properties of this system, for $m$-th eigenstate $|\Psi_m(k)\rangle$, we consider the Berry phase:
\begin{equation}
\Gamma_m =i\oint\langle\Psi_m(k)|\frac{d}{dk}|\Psi_m(k)\rangle dk,
\end{equation}
Here we need consider all the crystal momentum $k\in [-\frac{\pi}{L}, \frac{\pi}{L})$. The Berry phase of the ground state is $\tilde{\Gamma}=\sum_{m=1}^{L}\Gamma_m$ at half filling and we consider $\tilde{\gamma}=\tilde{\Gamma}$ mod $2\pi$, which equals to $\pi$ and $0$ for topological phase and trivial phase respectively. Fig.~\ref{S6} (b) displays the $\tilde{\gamma}$ as a function of $M_z$ and we see that the topological transition points satisfy $M_z=2(t_0+t_{so})$, which occurs concurrently with the critical-localization transition.

Corresponding to the nontrivial Berry phase, there should exist boundary states in the gap under OBC. As shown in Fig.~\ref{S6} (c),
there appear a pair of zero energy states when $M_z < 2(t_0+t_{so})$, which distribute near the boundaries, for example $M_z=1.5$, and the two zero energy states disappear when $M_z > 2(t_0+t_{so})$, for example $M_z=3.5$, and the corresponding eigenstates locate inside of the bulk.
Next we make analytical derivation of the topological transition points by using the transfer matrix approach~\cite{tma}. If introducing a local transformation
$\phi_{m,j}=u_{m,j}+v_{m,j}$ and $\psi_{m,j}=u_{m,j}-v_{m,j}$, from Eq.(\ref{tb3}) and Eq.(\ref{tb4}), we can obtain
\begin{equation}
 -t_0(\psi_{m,j+1}+\psi_{m,j-1})+\delta_j\psi_{m,j}+t_{so}(\psi_{m,j-1}-\psi_{m,j+1})=E_m\phi_{m,j},
\label{tb3n}
\end{equation}
\begin{equation}
 -t_0(\phi_{m,j+1}+\phi_{m,j-1})+\delta_j\phi_{m,j}+t_{so}(\phi_{m,j+1}-\phi_{m,j-1})=E_m\psi_{m,j}.
\label{tb4n}
\end{equation}
For zero energy states, the two equations can be represented in the transfer matrix form
\begin{equation}
 \begin{pmatrix}
 \psi_{j+1}\\
 \psi_j
\end{pmatrix}
=A_j
\begin{pmatrix}
 \psi_{j}\\
 \psi_{j-1}
\end{pmatrix}
\ \textrm{where} \ \
A_j=
\begin{pmatrix}
 \frac{\delta_j}{t_0+t_{so}} & \frac{t_{so}-t_0}{t_{so}+t_0}\\
1 & 0
\end{pmatrix}
%\notag
\end{equation}
A similar expression holds for the transfer matrix $B_j$ for the $\phi_j$.
If both the two eigenvalues of the full transfer matrix $\textsl{A}\equiv\Pi_{j=1}^LA_j$ are less than $1$ or larger than $1$, the system is topological (with a $\psi$ mode at one end of the chain)~\cite{tma}. We take $t_0=1$ and $t_{so}>0$, then the two eigenvalues of $\textsl{A}$ satisfy $|\lambda_1\lambda_2|<1$. Thus, $|\lambda_1| < 1$ if setting $|\lambda_1| < |\lambda_2|$, and the topological property is determined by the larger eigenvalue $|\lambda_2|$.

For $0< t_{so}< 1$, we perform a transformation $A_j=\sqrt{\Delta}S\tilde{A}_jS^{-1}$ with $S={\rm diag}(\Delta^{1/4},1/\Delta^{1/4})$ and $\Delta=\frac{1-t_{so}}{1+t_{so}}$. The matrices
$\tilde{A}_j=
\left(
\begin{matrix}
 \frac{\delta_j}{\sqrt{1-t_{so}^2}} & -1\\
 1 & 0
 \end{matrix}\right)$. Then one can obtain
\begin{equation}
 \textsl{A} (M_z, t_{so})=(\sqrt{\frac{1-t_{so}}{1+t_{so}}})^LS\textsl{A}(M_z/\sqrt{1-t_{so}^2}, 0)S^{-1}.
 \label{Amatrix}
\end{equation}
If introducing the Lyapunov exponent $\Upsilon\equiv \lim_{L\rightarrow\infty}\frac{1}{L}\ln|\lambda_2(M_z, t_{so})|$, which is the inverse of the localization length, from Eq.(\ref{Amatrix}), we can obtain
\begin{equation}
 \Upsilon(M_z, t_{so})=\Upsilon(\frac{M_z}{\sqrt{1-t_{so}^2}}, 0)-\frac{1}{2}\ln(\frac{1+t_{so}}{1-t_{so}}).
 \label{Lyp}
\end{equation}
When $t_{so}=0$, the model is reduced to the Aubry-Andr\'{e} model and the Lyapunov exponent $\Upsilon(M_z, 0)= \ln(M_z/2)$~\cite{AA}, so $\Upsilon(\frac{M_z}{\sqrt{1-t_{so}^2}}, 0)= \ln(\frac{M_z}{2\sqrt{1-t_{so}^2}})$. According to the above discussions, the topological transition point is at $|\lambda_2|=1$, i.e., $\Upsilon(M_z, t_{so})=0$. From Eq.(\ref{Lyp}), we obtain that the topological transition point obeys $M^c_z = 2(t_0+t_{so})$.
When $t_{so}>1$, by making the transformation $t_{so}\rightarrow 1/t_{so}$ and $\delta_{j}\rightarrow \delta_{j}/t_{so}$, we can obtain the same form of the phase boundary, i.e.,$M^c_z = 2(t_0+t_{so})$, which coincides with the critical-localization transition.

\section{V. Diffusion dynamics}
%%%%%%%%%%%%%%%%%%%%%%%%%%%%%%%%%%%%%%%%%%%%%%%
\begin{figure}
\centering
\includegraphics[width=0.45\textwidth]{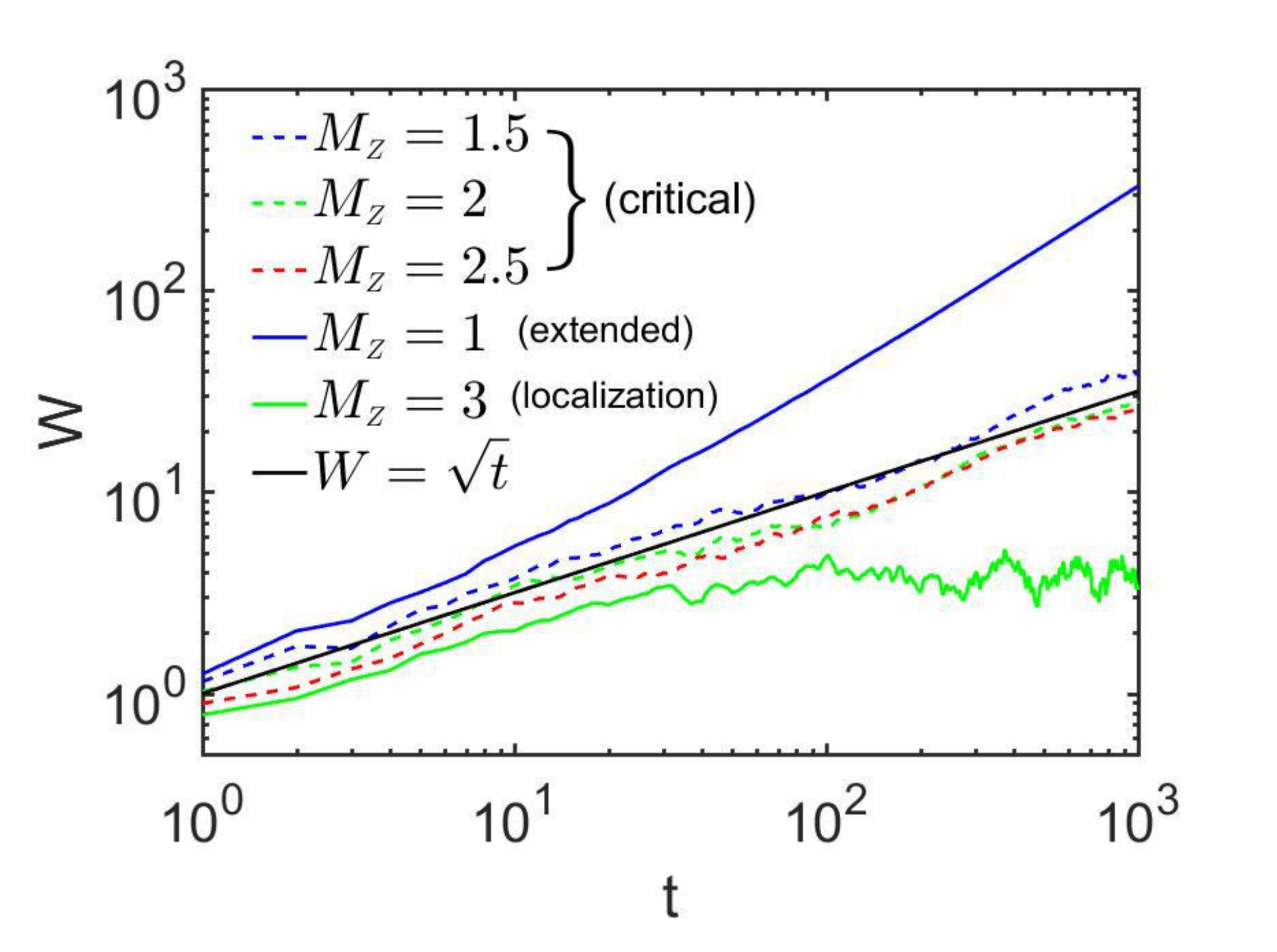}
\caption{\label{dyn}
Log-log plot of $W$ versus time $t$ for several values $M_z$ with fixed $t_0=1$, $t_{so}=0.3$ and $L=3000$. Here we take open boundary conditions.}
\end{figure}
%%%%%%%%%%%%%%%%%%%%%%%%%%%%%%%%%%%%%%%%%%%%%%%%
Below we quantitatively describe the differences of the dynamical evolutions for the three phases in the non-interacting case. We assume that a particle with spin down lie in the center at
$j=\frac{L}{2}$ with $L=3000$ at the time $t=0$. As mentioned in the main text, we consider the mean square displacement of the wave packet~\cite{Abe1988S,Qian}, which is given by
$W(t)=\sqrt{\sum_j(j-\frac{L}{2})^2(\langle n_{j,\uparrow}(t) + n_{j,\downarrow}(t)\rangle)}$. The time evolution of $W(t)$
can be expressed as $W(t)\sim t^{\kappa}$, where the dynamical index $\kappa$ takes three values for Aubry-Andr\'{e} model: $\kappa=1$, $\kappa=0$ and $\kappa\approx\frac{1}{2}$ in extended phase, localized phase and critical point respectively~\cite{Abe1988S,Qian}, indicating that the corresponding evolutions are ballistic, localized and approximately normal diffusive respectively. Fig. \ref{dyn} exhibits the result of $W$ versus $t$ for different $M_z$. When $M_z < 2|t_0-t_{so}|$ and $M_z > 2(t_0+t_{so})$, the wave packet motions are balllistic and localized respectively. In the case of $2|t_0-t_{so}| < M_z < 2(t_0+t_{so})$ where all the states are critical, the index $\kappa$ is approximate to $\frac{1}{2}$ independently of $M_z$, so the wave packet motion is approximately normal diffusion in whole critical region.

\begin{figure}
\includegraphics[width=0.90\textwidth]{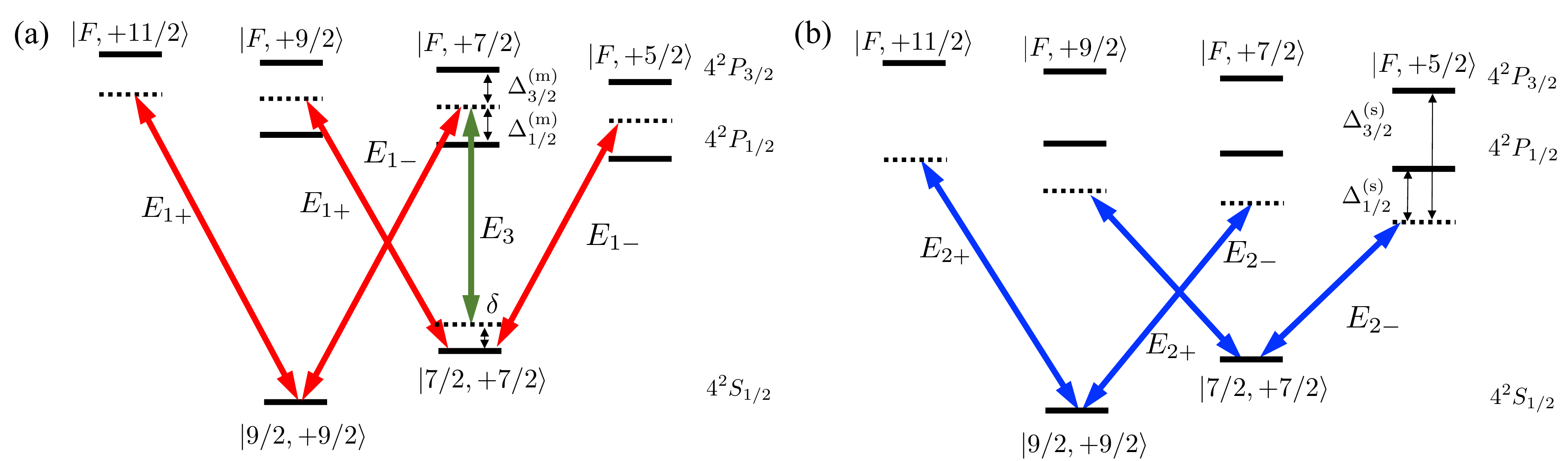}
\caption{The optical couplings for the realization. (a) Optical transitions for generating the primary lattice and Raman potentials ($^{40}$K atoms). (b) Optical transitions for generating incommensurate Zeeman potential.}\label{fig_levels}
\end{figure}

\section{VI. Experimental realization for $^{40}$K atoms}

As shown in Fig.~4 of the main text, we realize the model in a 1D optical Raman lattice with an incommensurate, spin-dependent weak lattice.
The total Hamiltonian reads
\begin{align}\label{Htotal}
H=\left[\frac{k_z^2}{2m}+V_1(z)\right]\otimes\id+{\cal M}(z)\sigma_x+\left[V_2(z)+\frac{\delta}{2}\right]\sigma_z,
\end{align}
where $V_1(z)$ denotes the primary lattice, $V_2(z)$ is the secondary spin-dependent weak lattice, giving an incommensurate Zeeman potential, ${\cal M}(z)$ is the Raman coupling potential, and $\delta$ is the two-photon detuning.
This setup makes use of two standing waves and one plane-wave beam. The standing-wave fields ${\bf E}_{1(2)}=\hat{e}_+ E_{1(2)+}(z)+\hat{e}_- E_{1(2)-}(z)$, where
\begin{eqnarray}
E_{1\pm}(z)&=&\sqrt{2} E_{1}e^{\ui(\phi_1+\phi_{1L}/2)}\cos(k_1 z-\phi_{1L}/2),\\
E_{2+}(z)&=&\sqrt{2}E_2e^{\ui(\phi_2+\phi_{2L}/2-\pi/4)}\cos(k_2z-\phi_{2L}/2+\pi/4), \\
E_{2-}(z)&=&\sqrt{2}E_2e^{\ui(\phi_2+\phi_{2L}/2+\pi/4)}\sin(k_2z-\phi_{2L}/2+\pi/4),
\end{eqnarray}
and the plane-wave field ${\bf E}_3=\hat{e}_zE_3 e^{\ui (k_3x+\phi_3)}$. Here $E_j$ ($j=1,2,3$) are the amplitudes, $\phi_j$ denote the initial phases, and $\phi_{1L}$ ($\phi_{2L}$) is the phase acquired by ${\bf E}_1$ (${\bf E}_2$) for an additional optical path back to the atom cloud.
In the following we take $^{40}$K atoms as an example while our results are generally applicable. For $^{40}$K, the spin-$1/2$ system can be constructed by $|\uparrow\rangle=|F=7/2, m_F=+7/2\rangle$ and $|\downarrow\rangle=|9/2, +9/2\rangle$. As shown in Fig.~\ref{fig_levels},
the lattice and Raman potentials are contributed from both the $D_2$ ($4{^{2}S}_{1/2}\to4{^{2}P}_{3/2}$) and $D_1$ ($4^{2}S_{1/2}\to4^{2}P_{1/2}$) lines.

The spin-independent main lattice $V_1(z)$ is created by the standing-wave field ${\bf E}_{1}$, given by
\begin{align}
V_{1\sigma}(z)=\sum_{F,\alpha=\pm}\frac{\left|\Omega^{(3/2)}_{\sigma F,1\alpha}\right|^2}{\Delta^{\rm (m)}_{3/2}}+\sum_{F,\alpha=\pm}\frac{\left|\Omega^{(1/2)}_{\sigma F,1\alpha}\right|^2}{\Delta^{\rm (m)}_{1/2}},
\end{align}
where $\Omega_{\sigma F,1+}^{(J)}=\langle\sigma|er|F,m_{F\sigma}+1,J\rangle\hat{e}_+\cdot{\bf E}_{1}$ and $\Omega_{\sigma F,1-}^{(J)}=\langle\sigma|er|F,m_{F\sigma}-1,J\rangle\hat{e}_-\cdot{\bf E}_{1}$ ($J=1/2,3/2$). From the dipole matrix elements of
$^{40}$K~\cite{Potassium}, we obtain
\begin{align}
V_{1}(z)=V_{\rm m}\cos^2(k_1 z-\phi_{1L}/2), \ V_{\rm m}=\frac{4t_{1/2}^2}{3}\left(\frac{2}{\Delta^{\rm (m)}_{3/2}}+\frac{1}{\Delta^{\rm (m)}_{1/2}}\right)E^2_{1},
\end{align}
with the transition matrix elements $t_{1/2}\equiv\langle J=1/2||e{\bf r}||J'=1/2\rangle$, $t_{3/2}\equiv\langle J=1/2||e{\bf r}||J'=3/2\rangle$ and $t_{3/2}\approx\sqrt{2}t_{1/2}$.
Similarly, the secondary, spin-dependent lattice is written as [see Fig.~\ref{fig_levels}(b)]
\begin{align}
V_{2\sigma}(z)=\sum_{F,\alpha=\pm}\frac{\left|\Omega^{(3/2)}_{\sigma F,2\alpha}\right|^2}{\Delta^{\rm (s)}_{3/2}}+\sum_{F,\alpha=\pm}\frac{\left|\Omega^{(1/2)}_{\sigma F,2\alpha}\right|^2}{\Delta^{\rm (s)}_{1/2}},
\end{align}
which (neglecting the small irrelevant spin-independent part) leads to the spin-dependent potential
\begin{align}
V_{2}(z)=V_{\rm s}\cos^2(k_2z-\phi_{2L}/2+\pi/4),\ V_{\rm s}=\frac{32t_{1/2}^2}{27}\left(\frac{1}{\Delta^{\rm (s)}_{1/2}}-\frac{1}{\Delta^{\rm (s)}_{3/2}}\right)E^2_{2}.
\end{align}
The Raman coupling potential ${\cal M}(z)$ is generated by the $E_{1+}$ component and the plane-wave light ${\bf E}_3$, given by
 \begin{align}
{\cal M}(z)=\sum_{F,\alpha=\pm}\frac{\Omega^{(3/2)*}_{\uparrow F,1z}\Omega^{(3/2)}_{\downarrow F,1-}}{\Delta^{\rm (m)}_{3/2}}
 +\sum_{F,\alpha=\pm}\frac{\Omega^{(1/2)*}_{\uparrow  F,1z}\Omega^{(1/2)}_{\downarrow F,1-}}{\Delta^{\rm (m)}_{1/2}},
 \end{align}
with $\Omega_{\uparrow F,1z}^{(J)}=\langle\uparrow|er|F,+7/2,J\rangle\hat{e}_z\cdot{\bf E}_{1}$,
 and takes the form ${\cal M}(z)=M_{\rm R}\cos(k_1 z-\phi_{1L}/2)e^{-\ui (k_3x+\phi_3-\phi_1-\phi_{1L}/2)}$. Hence, the Raman coupling strength reads
 \begin{align}
 M_{\rm R}=\frac{4\sqrt{2}t_{1/2}^2}{9}\left(\frac{1}{\Delta^{\rm (m)}_{3/2}}-\frac{1}{\Delta^{\rm (m)}_{1/2}}\right)E_1E_3.
 \end{align}
We set the wavelengths of ${\bf E}_{1,2}$ to be $\lambda_1=$768nm and $\lambda_2=$780nm, which leads to $\Delta^{\rm (m)}_{3/2}\simeq-\Delta^{\rm (m)}_{1/2}=\Delta_{\rm FS}/2$ and $\Delta^{\rm (s)}_{3/2}\simeq 4\Delta^{\rm (s)}_{1/2}/3=4\Delta_{\rm FS}$, where $\Delta_{\rm FS}$ denotes the fine structure splitting.
We then have $V_{\rm m}=8t_{1/2}^2 E_1^2/(3\Delta_{\rm FS})$, $V_{\rm s}=8t_{1/2}^2 E_2^2/(81\Delta_{\rm FS})$ and $M_{\rm R}=16\sqrt{2}t_{1/2}^2 E_1E_3/(9\Delta_{\rm FS})$, which yields $V_{\rm s}/V_{\rm m}\sim E_2^2/27E_1^2$ and $M_{\rm R}/V_{\rm m}\sim E_3/E_1$.

In the tight-binding regime and for $s$-bands, the Hamiltonian (\ref{Htotal}) takes the form of Eq.~(2) in the main text with
the spin-conserved and spin-flipped hopping coefficients $t_0$ and $t_{\rm so}$ respectively given by
\begin{align}
t_0=-\int dz\phi_{s}(z)\left[\frac{k_z^2}{2m}+V_{1}(z)\right]\phi_{s}(z-a),\ t_{\rm so}=M_{\rm R}\int dz\phi_{s}(z)\cos(k_1z)\phi_{s}(z-a),
\end{align}
and the incommensurate Zeeman potential
\begin{align}
\delta_j=\int dzV_{2}(z)|\phi_{s}(z-ja)|^2=\left[\frac{V_{\rm s}}{2}\int dz\cos(2k_2z)|\phi_{s}(z)|^2\right]\cos(2\pi\beta j)\equiv M_z\cos(2\pi\beta j).
\end{align}
Here $a=\pi/k_1$ is the lattice period, $\phi_{s}(z)$ denotes the Wannier function, and $\beta=k_2/k_1$.

%%%%%%%%%%%%%%%%%%%%%%%%%%%%%%%%%%%%%%%%%%%%%%

%%%%%%%%%%%%%%%%%%%%%%%%%%%%%%%%%%%%%%%%%%%%%


\begin{thebibliography}{36}
\bibitem{Anderson1958} P. W. Anderson, Absence of diffusion incertain random lattices, Phys. Rev. {\bf 109}, 1492 (1958).
\bibitem{Billy2008} J. Billy, V. Josse, Z. Zuo, A. Bernard, B. Hambrecht, P. Lugan, D. Cl\'{e}ment, L. Sanchez-Palencia, P. Bouyer, and A. Aspect, Direct observation of Anderson localization of matter waves in a controlled disorder, Nature (London) {\bf 453}, 891 (2008).
\bibitem{Roati2008} G. Roati, C. D'Errico, L. Fallani, M. Fattori, C. Fort, M. Zaccanti, G. Modugno, M. Modugno, and M. Inguscio, Anderson localization of a non-interacting Bose-Einstein condensate, Nature (London) {\bf 453}, 895 (2008).
\bibitem{Huse2010} A. Pal and D. A. Huse, Many-body localization phase transition, Phys. Rev. B {\bf 82}, 174411 (2010).
\bibitem{Huse2015} R. Nandkishore and D. A. Huse, Many-body localization and thermalization in quantum statistical mechanics, Annu. Rev. Condens. Matter Phys. {\bf 6}, 15 (2015).
\bibitem{Altman2015} E. Altman and R. Vosk, Universal dynamics and renormalization in many-body-localized systems, Annu. Rev. Condens. Matter Phys. {\bf 6}, 383 (2015).
\bibitem{HuseX} R. Vosk, D. A. Huse, and E. Altman, Theory of the many-body localization transition in one-dimensional systems, Phys. Rev. X {\bf 5}, 031032 (2015).
\bibitem{Li} X. Li, S. Ganeshan, J. H. Pixley, and S. D. Sarma, Many-body localization and quantum nonergodicit in a model with a single-particle mobility edge, Phys. Rev.
Lett. {\bf 115}, 186601 (2015).
\bibitem{Lea1} E. J. Torres-Herrera, and L. F. Santos, Dynamics at the many-body localization transition, Phys. Rev. B {\bf 92}, 014208 (2015).
\bibitem{Abanin2019} D. A. Abanin, E. Altman, I. Bloch, and M. Serbyn, Colloquium: Many-body localization, thermalization, and entanglement, Rev. Mod. Phys. 91, 021001 (2019).
\bibitem{Bloch1} M. Schreiber, S. S. Hodgman, P. Bordia, H. P. L\"{u}schen, M.
H. Fischer, R. Vosk, E. Altman, U. Schneider, and I. Bloch, Observation of many-body localization of interacting fermions in a quasirandom optical lattice,
Science {\bf 349}, 842 (2015).
\bibitem{Bordia} P. Bordia, H. P. L\"{u}schen, S. S. Hodgman, M. Schreiber, I.
Bloch, and U. Schneider, Coupling identical one-dimensional many-body localized systems, Phys. Rev. Lett. {\bf 116}, 140401 (2016).
\bibitem{Bloch2} P. Bordia, H. L\"{u}schen, S. Scherg, S. Gopalakrishnan, M.
Knap, U. Schneider, and I. Bloch, Probing slow relaxation and many-body localization in two-dimensional quasiperiodic systems, Phys. Rev. X {\bf 7}, 041047
(2017).
\bibitem{Bloch3} H. P. L\"{u}schen, P. Bordia, S. Scherg, F. Alet, E. Altman, U.
Schneider, and I. Bloch, Observation of slow dynamics near the many-body localization transition in one-dimensional quasiperiodic systems, Phys. Rev. Lett. {\bf 119}, 260401 (2017).
\bibitem{Bloch4} T. Kohlert, S. Scherg, X. Li, H. P. L\"{u}schen, S. D. Sarma, I. Bloch, and M. Aidelsburger, Observation of many-body localization in a one-dimensional system with a single-particle mobility edge, Phys. Rev. Lett. {\bf 122}, 170403 (2019).
\bibitem{Wang2019} Y. Wang, X.-J. Liu, and D. Yu, Many-body critical phase: extended and nonthermal, arXiv: 1910.12080.
\bibitem{Geisel1991} T. Geisel R. Ketzmerick and G. Petschel, New class of level statistics in quantum systems with unbounded diffusion,
Phys. Rev. Lett. {\bf 66},1651 (1991).
\bibitem{Fujita1986} K. Machida and M. Fujita, Quantum energy spectra and one-dimensional quasiperiodic systems,
Phys. Rev. B {\bf 34}, 7367 (1986).

\bibitem{GG2016} C. L. Bertrand and A. M. Garc\'{\i}a-Garc\'{\i}a, Anomalous thouless energy and critical statistics on the metallic side of the many-body localization transition, Phys. Rev. B {\bf 94}, 144201 (2016).
\bibitem{Halsey1986} T. C. Halsey, M. H. Jensen, L. P. Kadanoff, I. Procaccia, and B. I. Shraiman, Fractal measures and their singularities: The characterization of strange sets, Phys. Rev. A 33, 1141 (1986).
\bibitem{Mirlin2006} A. D. Mirlin, Y. V. Fyodorov, A. Mildenberger, and F. Evers, Exact relations between multifractal exponents at the Anderson transition,
  Phys. Rev. Lett. {\bf 97}, 046803 (2006).
\bibitem{Dubertrand2014} R. Dubertrand, I. Garc\'{\i}a-Mata, B. Georgeot, O. Giraud, G. Lemari\'{e}, and J. Martin, Two scenarios for quantum multifractality breakdown, Phys. Rev. Lett. {\bf 112}, 234101 (2014).
\bibitem{Abe1988} H. Hiramoto and S. Abe, Dynamics of an Electron in Quasiperiodic Systems. II. Harper's Model, J. Phys. Soc. Jpn. {\bf 57}, 1365 (1988).
\bibitem{Geisel1997} R. Ketzmerick, K. Kruse, S. Kraut, and T. Geisel, What determines the spreading of a wave packet?, Phys. Rev. Lett. {\bf 79}, 1959 (1997).
\bibitem{Modugno2009} M. Larcher, F. Dalfovo, and M. Modugno, Effects of interaction on the diffusion of atomic matter waves in one-dimensional quasiperiodic potentials, Phys. Rev. A {\bf 80}, 053606 (2009).

\bibitem{Hatsugai1990} Y. Hatsugai and M. Kohmoto, Energy spectrum and the quantum Hall effect on the square lattice with next-nearest-neighbor hopping, Phys. Rev. B {\bf 42}, 8282 (1990).
\bibitem{Takada2004} Y. Takada, K. Ino, and M. Yamanaka, Statistics of spectra for critical quantum chaos in one-dimensional quasiperiodic systems, Phys. Rev. E {\bf 70}, 066203 (2004).
\bibitem{Chong2015} F. Liu, S. Ghosh, and Y. D. Chong, Localization and adiabatic pumping in a generalized Aubry-Andr\'{e}-Harper model, Phys. Rev. B {\bf 91}, 014108 (2015).

\bibitem{Cai2013} X. Cai, L.-J. Lang, S. Chen, and Y. Wang, Topological superconductor to Anderson localization transition in one-dimensional incommensurate lattices, Phys. Rev. Lett. {\bf 110}, 176403 (2013).
\bibitem{Hu2016} J. Wang, X.-J. Liu, X. Gao, and H. Hu, Phase diagram of a non-Abelian Aubry-Andr\'{e}-Harper model with p-wave superfluidity, Phys. Rev. B {\bf 93}, 104504 (2016).
\bibitem{Wang2016} Y. Wang, Y. Wang, and S. Chen, Spectral statistics, finite-size scaling and multifractal analysis of quasiperiodic chain with p-wave pairing, Eur. Phys. J. B {\bf 89}, 254 (2016).

\bibitem{LiuXJ2013} X.-J. Liu, Z.-X. Liu, and M. Cheng, Manipulating topological edge spins in a one-dimensional optical lattice, Phys. Rev. Lett. {\bf 110}, 076401 (2013).
\bibitem{LiuXJ2014} X.-J. Liu, K. T. Law, and T. K. Ng, Realization of 2D spin-orbit interaction and exotic topological orders in cold atoms, Phys. Rev. Lett. {\bf 112}, 086401 (2014).
\bibitem{Pan2015} J.-S. Pan, X.-J. Liu, W. Zhang, W. Yi, and G.-C. Guo, Topological superradiant states in a degenerate Fermi gas, Phys. Rev. Lett. {\bf 115}, 045303 (2015).

\bibitem{Lepori2016} L. Lepori, I. C. Fulga, A. Trombettoni, and M. Burrello, Double Weyl points and Fermi arcs of topological semimetals in non-Abelian gauge potentials, Phys. Rev. A {\bf 94}, 053633 (2016).
\bibitem{Zhou2017PRL} X. Zhou, J. -S. Pan, Z.-X. Liu, W. Zhang, W. Yi, G. Chen, and S. Jia, Symmetry-protected topological states for interacting fermions in alkaline-earth-like atoms, Phys. Rev. Lett. {\bf 119}, 185701 (2017).
\bibitem{WangBZ2018} B.-Z. Wang, Y.-H. Lu, W. Sun, S. Chen, Y. Deng, and X.-J. Liu, Dirac-, Rashba-, and Weyl-type spin-orbit couplings: Toward experimental realization in ultracold atoms, Phys. Rev. A {\bf 97}, 011605(R) (2018).

\bibitem{Pu2019} C. Zhu, L. Chen, H. Hu, X. J. Liu, and H. Pu, Spin-exchange-induced exotic superfluids in a Bose-Fermi spinor mixture, Phys. Rev. A {\bf 100}, 031602(R) (2019).
\bibitem{Zhao2019NJP} H. Hu, I. I Satija, and E. Zhao, Chiral and counter-propagating Majorana fermions in a p-wave superconductor, New J. Phys. {\bf 21}, 123014 (2019).
\bibitem{Zheng2019} Z. Zheng, Z. Lin, D. -W. Zhang, S.-L. Zhu, and Z. D. Wang, Chiral magnetic effect in three-dimensional optical lattices, Phys. Rev. Research, {\bf 1}, 033102 (2019).
\bibitem{Lu2019} Y.-H. Lu, B.-Z. Wang, and X.-J. Liu, Realizing and detecting the fundamental Weyl semimetal phase, arXiv:1911.07169.
\bibitem{Liu2016} Z. Wu, L. Zhang, W. Sun, X.-T. Xu, B.-Z. Wang, S.-C. Ji, Y. Deng, S. Chen,
X.-J. Liu, and J.-W. Pan, Realization of two-dimensional spin-orbit coupling for Bose-Einstein condensates, Science {\bf 354}, 83 (2016).
\bibitem{Sun2018} W. Sun, B.-Z. Wang, X.-T. Xu, C.-R. Yi, L. Zhang, Z. Wu, Y. Deng, X.-J. Liu, S. Chen, and J.-W. Pan, Highly Controllable
and Robust 2D Spin-Orbit Coupling for Quantum Gases, Phys. Rev. Lett. {\bf 121}, 150401 (2018).
\bibitem{Song2018} B. Song, L. Zhang, C. He, T. F. J. Poon, E. Hajiyev, S. Zhang, X.-J. Liu, and G.-B. Jo, Observation of symmetry-protected topological band with ultracold fermions, Sci. Adv. {\bf 4}, eaao4748 (2018).
\bibitem{Song2019} B. Song, C. He, S. Niu, L. Zhang, Z. Ren, X.-J. Liu, and G.-B. Jo, Observation of nodal-line semimetal with ultracold fermions
in an optical lattice, Nat. Phys. {\bf 15}, 911 (2019).
\bibitem{RamanReview2018} L. Zhang and X.-J Liu, spin orbit coupling and topological phases for ultracold atoms. arXiv:1806.05628.

\bibitem{Kohmoto1983} M. Kohmoto, Metal-insulator transition and scaling for incommensurate systems. Phys. Rev. Lett {\bf 26}, 1198 (1983).

\bibitem{SM} See Supplemental Material for details.

\bibitem{Roots2017} P. Jurcevic, H. Shen, P. Hauke, C. Maier, T. Brydges, C. Hempel, B. P. Lanyon, M. Heyl, R. Blatt, and C. F. Roots, Direct observation of dynamical quantum phase transitions in an interacting many-body system, Phys. Rev. Lett. {\bf 119}, 080501 (2017).

\bibitem{measureRP} In experiment, the spatial configuration of initial and finial states can be measured by e.g. quantum gas microscopy. One can approximate the return probability $P(t)\approx N_i/N_z$ after repeating the initialization and measurement by $N_z\gg1$ number of times, each with evolution time $t$, where $N_i$ is the number of times that the final state at $t$ returns to the initial state.

\bibitem{Luitz} D. J. Luitz, N. Laflorencie, and F. Alet, Many-body localization edge in the random-field Heisenberg chain, Phys. Rev. B {\bf 91}, 081103 (2015).

\bibitem{Rigol2008} M. Rigol, V. Dunjko, and M. Olshanii, Thermalization and its mechanism for generic isolated quantum systems, Nature {\bf 452}, 854 (2008).

\bibitem{reason} Because the corresponding Hilbert space size ($D_H={{16}\choose{4}}=1820$) for $L=8$ isn't large enough and $\eta$ close to zero near the transition point between the MBC and MBL phases.

\bibitem{Mandel2003} O. Mandel, M. Greiner, A. Widera, T. Rom, T. W. H\"ansch, and I. Bloch, Coherent transport of neutral atoms in spin-dependent optical lattice potentials, Phys. Rev. Lett. {\bf 91}, 010407 (2003).

\end{thebibliography}

\begin{thebibliography}{99}
\bibitem{Hashimoto} Y. Hashimoto, K. Niizeki, and Y. Okabe, A finite-size scaling analysis of the localization properties of one-dimensional quasiperiodic systems, J. Phys. A {\bf 25}, 5211 (1992).
\bibitem{Wang2} Y. Wang, Y. Wang, and S. Chen, Spectral statistics, finite-size scaling and multifractal analysis of quasiperiodic chain with p-wave pairing, Eur. Phys. J. B {\bf 89}, 254 (2016).
\bibitem{Liu2013} X.-J. Liu, Z.-X. Liu, and M. Cheng, Manipulating topological edge spins in a one-dimensional optical lattice, Phys. Rev. Lett. {\bf 110}, 076401 (2013).
\bibitem{tma} W. DeGottardi, D. Sen, and S. Vishveshwara, Topological phases, Majorana modes and quench dynamics in a spin ladder system, New. J. Phys. {\bf 13}, 065028 (2011); W. DeGottardi, D. Sen, and S. Vishveshwara, Majorana fermions in superconducting 1D systems having periodic, quasiperiodic, and disordered potentials, Phys. Rev. Lett. {\bf 110}, 146404 (2013).
\bibitem{AA} S. Aubry and G. Andr\'{e}, Analyticity breaking and Anderson localization in incommensurate lattices, Ann. Isr. Phys. Soc. {\bf 3}, 133 (1980).
\bibitem{Abe1988S} H. Hiramoto and S. Abe, Dynamics of an Electron in Quasiperiodic Systems. II. Harper's Model, J. Phys. Soc. Jpn. {\bf 57}, 1365 (1988).
\bibitem{Qian} J. Zhong, R. B. Diener, D. A. Steck, W. H. Oskay, M. G. Raizen,
E. W. Plummer, Z. Zhang, and Q. Niu, Shape of the quantum diffusion front, Phys. Rev. Lett {\bf 86}, 2485 (2001).
\bibitem{Potassium} T. G. Tiecke, Properties of Potassium, 2011, http://www.tobiastiecke.nl/archive/PotassiumProperties.pdf.
\end{thebibliography}
\end{document}